\documentclass[11pt,a4paper]{article}

\usepackage{amssymb}
\usepackage[dvips]{graphicx}
\usepackage{bm}

\begin{document}

\title{\bf Three-loop contribution of the Faddeev--Popov ghosts to the $\beta$-function of ${\cal N}=1$ supersymmetric gauge theories and the NSVZ relation}

\author{
\smallskip
M.D.Kuzmichev, N.P.Meshcheriakov, S.V.Novgorodtsev,\\
\smallskip
I.E.Shirokov, K.V.Stepanyantz\\
{\small{\em Moscow State University}}, {\small{\em  Faculty of Physics,}}\\
{\small{\em Department  of Theoretical Physics,}}\\
{\small{\em 119991, Moscow, Russia}}}

\maketitle

\begin{abstract}
We find the three-loop contribution to the $\beta$-function of ${\cal N}=1$ supersymmetric gauge theories regularized by higher covariant derivatives produced by the supergraphs containing loops of the Faddeev--Popov ghosts. This is done using a recently proposed algorithm, which essentially simplifies such multiloop calculations. The result is presented in the form of an integral of double total derivatives in the momentum space. The considered contribution to the $\beta$-function is compared with the two-loop anomalous dimension of the Faddeev--Popov ghosts. This allows verifying the validity of the NSVZ equation written as a relation between the $\beta$-function and the anomalous dimensions of the quantum superfields. It is demonstrated that in the considered approximation the NSVZ equation is satisfied for the renormalization group functions defined in terms of the bare couplings. The necessity of the nonlinear renormalization for the quantum gauge superfield is also confirmed.
\end{abstract}

\unitlength=1cm

\section{Introduction}
\hspace*{\parindent}

${\cal N}=1$ supersymmetric models have better ultraviolet behaviour in comparison with the non-supersymmetric theories. Namely, the superpotential has no divergent quantum corrections \cite{Grisaru:1979wc}, the three-point vertices with two ghost legs and one leg of the quantum gauge superfield are finite \cite{Stepanyantz:2016gtk}, and the $\beta$-function is related to the anomalous dimension of the matter superfields $(\gamma_\phi)_i{}^j$ by the NSVZ equation \cite{Novikov:1983uc,Jones:1983ip,Novikov:1985rd,Shifman:1986zi}

\begin{equation}\label{NSVZ_Original_Form}
\frac{\beta(\alpha,\lambda)}{\alpha^2} = - \frac{3 C_2 - T(R) + C(R)_i{}^j \big(\gamma_\phi\big)_j{}^i(\alpha,\lambda)/r}{2\pi(1- C_2\alpha/2\pi)}
\end{equation}

\noindent
(see also \cite{Shifman:1999mv,Shifman:1999kf,Shifman:2018yxh}). In our notation $r$ and $f^{ABC}$ are the dimension and the structure constants of a simple gauge group $G$, respectively, and $f^{ACD} f^{BCD} \equiv C_2 \delta^{AB}$. The generators of the fundamental representation $t^A$ are normalized by the  condition $\mbox{tr}(t^A t^B) = \delta^{AB}/2$, while the generators of the representation $R$ to which the matter superfields belong satisfy the equations

\begin{equation}
\mbox{tr}(T^A T^B) = T(R)\delta^{AB};\qquad (T^A)_i{}^k (T^A)_k{}^j \equiv C(R)_i{}^j.
\end{equation}

The NSVZ $\beta$-function can be used for proving the finiteness of ${\cal N}=2$ supersymmetric gauge theories beyond the one-loop approximation \cite{Shifman:1999mv,Buchbinder:2014wra} in the case of a manifestly ${\cal N}=2$ supersymmetric quantization \cite{Galperin:1985ec,Galperin:1984av,Galperin:2001uw,Buchbinder:2001wy}, which in particular should include the invariant regularization \cite{Buchbinder:2015eva}. (Earlier, the ${\cal N}=2$ non-renormalization theorem has been obtained by different methods, see Refs. \cite{Grisaru:1982zh,Howe:1983sr,Buchbinder:1997ib}.) Consequently, the NSVZ relation also leads to the finiteness of ${\cal N}=4$ supersymmetric Yang--Mills (SYM) theory proved in \cite{Grisaru:1982zh,Howe:1983sr,Mandelstam:1982cb,Brink:1982pd} after the explicit three-loop calculation of Ref. \cite{Avdeev:1980bh}.

Using the non-renormalization theorem for the triple gauge-ghost vertices one can rewrite the NSVZ equation in the form of the relation between the $\beta$-function and the anomalous dimensions of the quantum gauge superfield, of the Faddeev--Popov ghosts, and of the matter superfields denoted by $\gamma_V$, $\gamma_c$, and $(\gamma_\phi)_i{}^j$, respectively,

\begin{equation}\label{NSVZ_Equation_New_Form}
\frac{\beta(\alpha,\lambda)}{\alpha^2} = - \frac{1}{2\pi} \Big(3 C_2 - T(R) - 2 C_2 \gamma_c(\alpha,\lambda) - 2 C_2 \gamma_V(\alpha,\lambda) + C(R)_i{}^j \big(\gamma_\phi\big)_j{}^i(\alpha,\lambda)/r \Big).
\end{equation}

The NSVZ relations (\ref{NSVZ_Original_Form}) and (\ref{NSVZ_Equation_New_Form}) are valid only for certain (NSVZ) renormalization prescriptions. Using the general equations describing how Eq. (\ref{NSVZ_Original_Form}) changes under finite renormalizations, it is possible to demonstrate that the NSVZ schemes form a continuous set \cite{Goriachuk:2018cac}. The $\overline{\mbox{DR}}$ scheme does not enter this set \cite{Jack:1996vg,Jack:1996cn,Jack:1998uj,Harlander:2006xq}. However, with the higher covariant derivative regularization \cite{Slavnov:1971aw,Slavnov:1972sq,Slavnov:1977zf} (see \cite{Krivoshchekov:1978xg,West:1985jx} for its various ${\cal N}=1$ supersymmetric versions) minimal subtractions of logarithms produce the NSVZ scheme in all orders at least in the Abelian case \cite{Kataev:2013eta,Kataev:2013csa,Kataev:2014gxa}. We call this renormalization prescription HD+MSL. The HD+MSL scheme seems to be NSVZ also in the non-Abelian case \cite{Stepanyantz:2016gtk}.  This implies that the higher covariant derivative regularization is much more convenient for making calculations in supersymmetric theories in comparison with the dimensional reduction \cite{Siegel:1979wq}.

An interesting feature of using the higher covariant derivative method for regularizing ${\cal N}=1$ supersymmetric gauge theories is that all integrals giving the $\beta$-function appear to be integrals of double total derivatives in the momentum representation. Due to this structure of the loop integrals all higher order ($L\ge 2$) contributions to the $\beta$-function originate from $\delta$-singularities. The factorization in total and double total derivatives with respect to the loop momenta was first noted in calculating the lowest quantum corrections in ${\cal N}=1$ SQED in Refs. \cite{Soloshenko:2003nc} and \cite{Smilga:2004zr}, respectively. The rigorous proof of this fact for ${\cal N}=1$ SQED has been done in \cite{Stepanyantz:2011jy,Stepanyantz:2014ima}. It also turns out that due to the factorization of the loop integrals into integrals of double total derivatives the NSVZ relation in ${\cal N}=1$ SQED is valid in the on-shell scheme in all loops \cite{Kataev:2019olb}.

The method developed in Ref. \cite{Stepanyantz:2011jy} has also been applied for constructing the all-loop proof of the NSVZ-like equation \cite{Hisano:1997ua,Jack:1997pa,Avdeev:1997vx} and deriving the NSVZ-like scheme for the renormalization of the photino mass in SQED with softly broken supersymmetry \cite{Nartsev:2016nym,Nartsev:2016mvn}. Also it works for the Adler $D$-function in ${\cal N}=1$ SQCD \cite{Shifman:2014cya,Shifman:2015doa,Kataev:2017qvk}.\footnote{Again, in the $\overline{\mbox{DR}}$-scheme the NSVZ-like equations are not valid, see Refs. \cite{Jack:1998uj} and \cite{Aleshin:2019yqj}.} However, the derivation of the NSVZ relation in the non-Abelian case turns out to be more complicated. Nevertheless, the calculations made with the higher covariant derivative regularization in the lowest orders (see, e.g., \cite{Pimenov:2009hv,Stepanyantz:2011cpt,Stepanyantz:2011bz,Shakhmanov:2017soc,Kazantsev:2018nbl}) reveal the same features as in the Abelian case. In particular, they demonstrate that all integrals for the $\beta$-function defined in terms of the bare couplings are integrals of double total derivatives. This allows to outline the following main steps for the perturbative all-loop derivation of the NSVZ equation:

1. Rewriting Eq. (\ref{NSVZ_Original_Form}) in the form (\ref{NSVZ_Equation_New_Form});

2. Proving the factorization of integrals giving the $\beta$-function into integrals of double total derivatives and reducing them to integrals of $\delta$-singularities;

3. Calculating the sum of the singular contributions.

Certainly, use of the higher covariant derivative regularization is very important and is always assumed. The first and second steps have been done in Refs. \cite{Stepanyantz:2016gtk} and \cite{Stepanyantz:2019ihw}, respectively. However, the singular contributions have not yet been summed. Presumably, the result should coincide with the terms containing the anomalous dimensions in Eq. (\ref{NSVZ_Equation_New_Form}). If it is so, then the NSVZ equation is satisfied by the renormalization group functions (RGFs) defined in terms of the bare couplings independently of the renormalization prescription which supplements the higher covariant derivative regularization. Consequently, for RGFs (standardly) defined in terms of the renormalized couplings the HD+MSL scheme appears to be NSVZ in all orders \cite{Stepanyantz:2016gtk}.

Although the sum of singularities has not yet been found, some calculations in the lowest orders indicate that it really gives the anomalous dimensions in the right hand side of Eq. (\ref{NSVZ_Equation_New_Form}). It was verified to the order $O(\alpha)$ inclusive in Ref. \cite{Shakhmanov:2017wji}, but in this approximation all terms of the NSVZ equation are scheme-independent.\footnote{The calculation of Ref. \cite{Shakhmanov:2017wji} has been done with a simplified version of the higher covariant derivative regularization which breaks the BRST invariance and, for this reason, supplemented by a special renormalization procedure \cite{Slavnov:2003cx} restoring the Slavnov--Taylor identities.} In the next order $O(\alpha^2)$ this has been done in Refs. \cite{Shakhmanov:2017soc,Kazantsev:2018nbl} for terms containing the Yukawa couplings. In the right hand side of Eq. (\ref{NSVZ_Equation_New_Form}) such terms are present inside the anomalous dimensions $(\gamma_\phi)_i{}^j$ and $\gamma_V$. However, at present no nontrivial verifications of the term containing $\gamma_c$ have been done. This is the purpose of the present paper. Namely, we obtain expressions for all three-loop contributions to the $\beta$-function coming from supergraphs containing the Faddeev--Popov ghost loops. Then, we extract terms which correspond to the cuts of internal ghost lines and compare them with the two-loop ghost anomalous dimension calculated in Ref. \cite{Kazantsev:2018kjx}.

To calculate the three-loop contributions to the $\beta$-function coming from the considered class of supergraphs, we will use the algorithm proposed in Ref. \cite{Stepanyantz:2019ihw}. It essentially simplifies the calculations and produces the result in the form of integrals of double total derivatives. Therefore, if we manage to obtain the $\gamma_c$ term in Eq. (\ref{NSVZ_Equation_New_Form}), this algorithm will also be tested by a highly nontrivial calculation.

The paper is organized as follows. In Sect. \ref{Section_SYM_HD} we briefly recall the main information about the application of the higher covariant derivative method for regularizing ${\cal N}=1$ supersymmetric gauge theories. The algorithm for constructing integrals of double total derivatives proposed in Ref. \cite{Stepanyantz:2019ihw} is described in Sect. \ref{Section_Algorithm}. The three-loop contribution to the $\beta$-function defined in terms of the bare couplings is calculated and compared with the two-loop ghost anomalous dimension in Sect. \ref{Section_Beta}.

\section{The higher covariant derivative regularization for ${\cal N}=1$ SYM theories with matter}
\hspace*{\parindent}\label{Section_SYM_HD}

In this paper we consider the ${\cal N}=1$ SYM theory interacting with chiral matter superfields in a certain representation $R$ of the gauge group $G$. It is convenient to formulate and quantize this theory in terms of ${\cal N}=1$ superfields, because in this case the calculation of quantum corrections is made in a manifestly ${\cal N}=1$ supersymmetric way. In this formulation the classical action in the massless case is written as

\begin{eqnarray}\label{Action_Of_Theory_Classical}
&& S_{\mbox{\scriptsize classical}} = \frac{1}{2 e_0^2}\,\mbox{Re}\,\mbox{tr}\int d^4x\,d^2\theta\, W^a W_a + \frac{1}{4} \int d^4x\,d^4\theta\,
\phi^{*i} (e^{2V})_i{}^j \phi_j\nonumber\\
&&\qquad\qquad\qquad\qquad\qquad\qquad\qquad\qquad\qquad +
\Big(\frac{1}{6} \lambda_0^{ijk} \int d^4x\,d^2\theta\, \phi_i \phi_j \phi_k + \mbox{c.c.}\Big),\qquad
\end{eqnarray}

\noindent
where the gauge superfield is denoted by $V$ and $\phi_i$ are chiral matter superfields. In this paper we adopt the notations, in which $V = e_0 V^A t^A$ inside the gauge superfield strength $W_a \equiv \bar D^2 (e^{-2V} D_a e^{2V})/8$ and $V= e_0 V^A T^A$ in the matter part of the action.

However, at the quantum level the gauge superfield is renormalized in a nonlinear way \cite{Piguet:1981fb,Piguet:1981hh,Tyutin:1983rg}. In the lowest-order approximation this nonlinear renormalization was found in Refs. \cite{Juer:1982fb,Juer:1982mp}. Another calculation made in Ref. \cite{Kazantsev:2018kjx} explicitly demonstrates that the renormalization group equations are satisfied only if this nonlinear renormalization is taken into account. That is why constructing the generating functional one should replace the gauge superfield $V$ by a nonlinear function

\begin{equation}\label{Y_Definition}
{\cal F}(V) = e_0 \Big(V^A + e_0^2 y_0\, G^{ABCD}\, V^B\, V^C\, V^D +\ldots \Big)\, t^A,
\end{equation}

\noindent
which includes an infinite set of parameters $Y_0=(y_0,\ldots)$ needed for performing the nonlinear renormalization. Here the coefficient $G^{ABCD}$ is a totally symmetric tensor defined as

\begin{equation}
G^{ABCD} \equiv \frac{1}{6}\Big(f^{AKL} f^{BLM} f^{CMN} f^{DNK} + \mbox{permutations of $B$, $C$, and $D$} \Big).
\end{equation}

Also we will use the background field method \cite{DeWitt:1965jb,Abbott:1980hw,Abbott:1981ke} formulated in terms of ${\cal N}=1$ superfields \cite{Grisaru:1982zh,Gates:1983nr}. Taking into account the necessity of introducing the function ${\cal F}(V)$, the quantum-background splitting is made with the help of the replacement

\begin{equation}
e^{2{\cal F}(V)} \to e^{2{\cal F}(V)} e^{2\bm{V}},
\end{equation}

\noindent
where $\bm{V}$ is the Hermitian background gauge superfield, and the quantum gauge superfield $V$ is restricted by the constraint $V^+ = e^{-2\bm{V}} V e^{2\bm{V}}$.

To regularize the theory under consideration, we add to the action certain terms containing the higher degrees of the supersymmetric covariant derivatives

\begin{equation}\label{Definition_Of_Covariant_Derivatives}
\nabla_a = D_a;\qquad \bar\nabla_{\dot a} = e^{2{\cal F}(V)} e^{2\bm{V}} \bar D_{\dot a} e^{-2\bm{V}} e^{-2{\cal F}(V)}.
\end{equation}

\noindent
Then the regularized action can be presented in the form

\begin{eqnarray}\label{Action_With_Regularization}
&& S_{\mbox{\scriptsize reg}} = \frac{1}{2 e_0^2}\,\mbox{Re}\,\mbox{tr} \int d^4x\, d^2\theta\, W^a \Big(e^{-2\bm{V}} e^{-2{\cal F}(V)}\Big)_{Adj} R\Big(-\frac{\bar\nabla^2 \nabla^2}{16\Lambda^2}\Big)_{Adj}\Big(e^{2{\cal F}(V)}e^{2\bm{V}}\Big)_{Adj} W_a\qquad\nonumber\\
&& + \frac{1}{4} \int d^4x\,d^4\theta\, \phi^{*i} \Big(F\Big(-\frac{\bar\nabla^2 \nabla^2}{16\Lambda^2}\Big) e^{2{\cal F}(V)}e^{2\bm{V}}\Big)_i{}^j \phi_j
+ \frac{1}{6} \Big(\lambda_0^{ijk} \int d^4x\,d^2\theta\, \phi_i \phi_j \phi_k + \mbox{c.c.} \Big),\qquad
\end{eqnarray}

\noindent
where the gauge superfield strength is defined as

\begin{equation}\label{Definition_Of_W}
W_a \equiv \frac{1}{8} \bar D^2 \left(e^{-2\bm{V}} e^{-2{\cal F}(V)}\, D_a \left(e^{2{\cal F}(V)}e^{2\bm{V}}\right)\right),
\end{equation}

\noindent
and we use the notation

\begin{equation}
\Big(f_0 + f_1 X + f_2 X^2 + \ldots\Big)_{Adj} Y = f_0 Y + f_1 [X, Y] + f_2 [X,[X,Y]] + \ldots
\end{equation}

\noindent
The functions $R(x)$ and $F(x)$, such that $R(0)=F(0)=1$, should rapidly increase at infinity. (This allows to provide the finiteness of the regularized superdiagrams beyond the one-loop approximation.)

Following Refs. \cite{Kazantsev:2018kjx}, we will use the gauge fixing action

\begin{equation}\label{Gauge_Fixing}
S_{\mbox{\scriptsize gf}} = -\frac{1}{16\xi_0 e_0^2}\, \mbox{tr} \int d^8x\,  \bm{\nabla}^2 V R\Big(-\frac{\bm{\bar\nabla}^2 \bm{\nabla}^2}{16\Lambda^2}\Big)_{Adj} \bm{\bar\nabla}^2 V,
\end{equation}

\noindent
where the supersymmetric background covariant derivatives are defined as

\begin{equation}\label{Definition_Of_Background_Derivatives}
\bm{\nabla}_a = D_a;\qquad \bm{\bar\nabla}_{\dot a} = e^{2\bm{V}} \bar D_{\dot a} e^{-2\bm{V}}.
\end{equation}

\noindent
Note that the expression (\ref{Gauge_Fixing}) is invariant under the background gauge transformations and contains the same regulator function $R$ as in Eq. (\ref{Action_With_Regularization}).

The Faddeev--Popov action corresponding to the gauge fixing term (\ref{Gauge_Fixing}) is written as

\begin{eqnarray}\label{Action_Ghosts}
&& S_{\mbox{\scriptsize FP}} = \frac{1}{2} \int d^4x\, d^4\theta\,
\frac{\partial {\cal F}^{-1}(\widetilde V)^A}{\partial {\widetilde V}^B}\left.\vphantom{\frac{1}{2}}\right|_{\widetilde V = {\cal F}(V)} \left(\big(e^{2\bm{V}}\big)_{Adj}\bar c +
\bar c^+ \right)^A\nonumber\\
&&\qquad\qquad\qquad\qquad \times \left\{\vphantom{\frac{1}{2}}\smash{\Big(\frac{{\cal F}(V)}{1-e^{2{\cal F}(V)}}\Big)_{Adj} c^+
+ \Big(\frac{{\cal F}(V)}{1-e^{-2{\cal F}(V)}}\Big)_{Adj}
\Big(\big(e^{2\bm{V}}\big)_{Adj} c \Big)}\right\}^B,\qquad
\end{eqnarray}

\noindent
where $c = e_0 c^A t^A$ and $\bar c = e_0 \bar c^A t^A$ are the chiral ghost and antighost superfields, respectively.

The Nielsen--Kallosh ghosts should also be introduced. However, they are essential only in the one-loop approximation, which was considered in Ref. \cite{Aleshin:2016yvj} in detail. That is why here we will not discuss them.

It should be mentioned that to regularize the one-loop divergences, one needs to insert the Pauli--Villars determinants into the generating functional \cite{Slavnov:1977zf,Faddeev:1980be}. Here, following Ref. \cite{Kazantsev:2017fdc}, we use three commuting chiral superfields $\varphi_1$, $\varphi_2$, and $\varphi_3$  in the adjoint representation and the chiral superfields $\Phi_i$ in a representation $R_{\mbox{\scriptsize PV}}$ for which it is possible to write the invariant mass term, such that $M^{ij} M^*_{jk} = M^2 \delta^i_k$. The former superfields cancel one-loop divergences coming from the gauge and ghost loops, while the latter ones cancel one-loop divergences introduced by the matter loop. The actions for the Pauli--Villars superfields and the explicit expression for the generating functional can be found in \cite{Stepanyantz:2019ihw}. It is important that the masses of the Pauli--Villars superfields should be proportional to the parameter $\Lambda$ in the higher derivative term,

\begin{equation}\label{Pauli_Villars_Masses}
M_\varphi = a_\varphi \Lambda;\qquad M = a\Lambda,
\end{equation}

\noindent
where $a_\varphi$ and $a$ are the coefficients independent of couplings.

We will define the renormalization constants by the equations

\begin{eqnarray}
&& \frac{1}{\alpha_0} = \frac{Z_\alpha}{\alpha};\qquad\qquad\quad\ \frac{1}{\xi_0} = \frac{Z_\xi}{\xi}; \qquad\qquad\quad\ \, \phi_i = \big(\sqrt{Z_\phi}\big)_i{}^j \big(\phi_R\big)_j;\qquad\nonumber\\
&& V = Z_V Z_\alpha^{-1/2} V_R;\qquad\ \bar c c = Z_c Z_\alpha^{-1} \bar c_R c_R;\qquad  y_0 = Z_y y,
\end{eqnarray}

\noindent
where $\alpha_0 = e_0^2/4\pi$ and $\alpha = e^2/4\pi$. The bare couplings are denoted by the subscript $0$, while the subscript $R$ denotes renormalized superfields. Note that in the considered approximation the nonlinear renormalization of the quantum gauge superfield corresponds to the renormalization of the parameter $y_0$ which was introduced in Eq. (\ref{Y_Definition}). According to Refs. \cite{Juer:1982fb,Juer:1982mp} and \cite{Kazantsev:2018kjx}, in the lowest-order approximation it can be written as

\begin{equation}\label{Renormalization_Of_Y}
y_0 = y + \frac{\alpha}{90\pi} \Big((2+3\xi) \ln\frac{\Lambda}{\mu} + k_1\Big) + \ldots,
\end{equation}

\noindent
where dots denote the higher order terms and $k_1$ is a finite constant.

In terms of the bare couplings RGFs are defined by the equations

\begin{eqnarray}
&&\hspace*{-5mm} \beta(\alpha_0,\lambda_0,Y_0) \equiv \left.\frac{d\alpha_0}{d\ln\Lambda}\right|_{\alpha,\lambda,Y = \mbox{\scriptsize const}};\qquad\quad \ \
\gamma_V(\alpha_0,\lambda_0,Y_0) \equiv \left. - \frac{d\ln Z_V}{d\ln\Lambda}\right|_{\alpha,\lambda,Y = \mbox{\scriptsize const}};\nonumber\\
&&\hspace*{-5mm} \gamma_c(\alpha_0,\lambda_0,Y_0) \equiv \left. - \frac{d\ln Z_c}{d\ln\Lambda}\right|_{\alpha,\lambda,Y = \mbox{\scriptsize const}};\qquad
(\gamma_\phi)_i{}^j(\alpha_0,\lambda_0,Y_0) \equiv \left. - \frac{d(\ln Z_\phi)_i{}^j}{d\ln\Lambda}\right|_{\alpha,\lambda,Y = \mbox{\scriptsize const}}.\qquad
\end{eqnarray}

\section{A method for calculating multiloop contributions to the $\beta$-function}
\hspace*{\parindent}\label{Section_Algorithm}

According to Ref. \cite{Stepanyantz:2019ihw}, it is possible to construct integrals of double total derivatives contributing to the $\beta$-function (defined in terms of the bare couplings in the case of using the higher covariant derivative regularization) with the help of a special algorithm, which essentially simplifies the calculations. This occurs, because this algorithm reduces the calculation of superdiagrams with two external lines of the background gauge superfield to the evaluation of certain superdiagrams without external lines. If we consider an $L$-loop supergraph without external lines, then the corresponding superdiagrams contributing to the two-point Green function of the background gauge superfield

\begin{equation}
\Gamma^{(2)}_{\bm{V}} = - \frac{1}{8\pi} \mbox{tr} \int \frac{d^4p}{(2\pi)^4}\, d^4\theta\, \bm{V}(-p,\theta) \partial^2 \Pi_{1/2} \bm{V}(p,\theta)\, d^{-1}(\alpha_0,\lambda_0,Y_0,\Lambda/p)
\end{equation}

\noindent
are obtained by attaching to it two external $\bm{V}$-lines in all possible ways. The function $d^{-1}$ is related to the $\beta$-function defined in terms of the bare couplings by the equation

\begin{equation}\label{Beta_Bare_Definition}
\frac{\beta(\alpha_0,\lambda_0,Y_0)}{\alpha_0^2} =\left. - \frac{d}{d\ln\Lambda}\Big(\frac{1}{\alpha_0}\Big)\right|_{\alpha,\lambda,Y = \mbox{\scriptsize const}} = \left.\frac{d}{d\ln\Lambda}\big(d^{-1} - \alpha_0^{-1}\big)\right|_{\alpha,\lambda,Y = \mbox{\scriptsize const};\ p = 0},
\end{equation}

\noindent
where $p$ is the external momentum, and the derivatives with respect to $\ln\Lambda$ should be calculated at fixed values of renormalized couplings. Thus, a supergraph without external lines can be matched to a certain contribution to the function $\beta/\alpha_0^2$. Various calculations made with the higher covariant derivative regularization demonstrate that such contributions are given by integrals of double total derivatives. According to Ref. \cite{Stepanyantz:2019ihw}, a contribution to $\beta/\alpha_0^2$ corresponding to a certain supergraph can be found with the help of a special algorithm, which is described below.

Let us consider an $L$-loop supergraph without external legs. Then, to obtain the corresponding contribution to $\beta/\alpha_0^2$, it is necessary to do the following:

1. Formally construct the corresponding contribution to the effective action using the superspace Feynman rules.

2. Formally insert the factor $\theta^4 (v^B)^2$ at an arbitrary point of the supergraph containing the integration over the full superspace,\footnote{If such points are absent, then it is necessary to convert one of the integrations over $d^4x\,d^2\theta$ into the integration over the full superspace.} where the slowly decreasing functions $v^B$ should tend to 0 at a sufficiently large scale $R\to \infty$. For example, it is possible to choose

\begin{equation}
v^B = v_0^B \exp\Big(-\big(X^\mu\big)^2/2R^2\Big),
\end{equation}

\noindent
where the Euclidean coordinates are denoted by $X^\mu = (x^i,\, ix^0)$ and $v_0^B = \mbox{const}$.

3. Calculate the resulting expression using the standard $D$-algebra and omit terms suppressed by powers of $1/(\Lambda R)$. Note that the limit $R\to\infty$ corresponds to the limit $p\to 0$ in Eq. (\ref{Beta_Bare_Definition}), and the result is always proportional to

\begin{equation}
{\cal V}_4 \equiv \int d^4x\, \big(v^B\big)^2 \to \infty.
\end{equation}

\noindent
(The functions $v^B$ are introduced in order to make ${\cal V}_4$ finite and to avoid dealing with expressions which are not well-defined.)

4. Mark $L$ propagators with independent (Euclidean) momenta $Q_i^\mu$ and the indices $a_i$ corresponding to their beginnings.

5. In the integrand of the loop integral formally replace the product

\begin{equation}\label{Product}
\prod\limits_{i=1}^L \delta_{a_i}^{b_i}
\end{equation}

\noindent
coming from the marked propagators by the differential operator

\begin{eqnarray}
\sum_{k,l=1}^L  \prod\limits_{i\ne k,l} \delta_{a_i}^{b_i}\, (T^A)_{a_k}{}^{b_k} (T^A)_{a_l}{}^{b_l} \frac{\partial^2}{\partial Q^\mu_k \partial Q^\mu_l}.
\end{eqnarray}

6. Multiply the result by the factor $-2\pi/(r{\cal V}_4)\cdot d/d\ln\Lambda$.

According to Ref. \cite{Stepanyantz:2019ihw}, the final expression gives a part of the expression

\begin{equation}
\frac{1}{\alpha_0^2}\Big(\beta(\alpha_0,\lambda_0, Y_0) -\beta_{\mbox{\scriptsize 1-loop}}(\alpha_0) \Big)
\end{equation}

\noindent
corresponding to the considered supergraph.

\section{Three-loop contribution to the $\beta$-function produced by ghost loops}
\hspace*{\parindent}\label{Section_Beta}

\begin{figure}[h]
\begin{picture}(0,8.4)
\put(1.2,5.8){\includegraphics[scale=0.2]{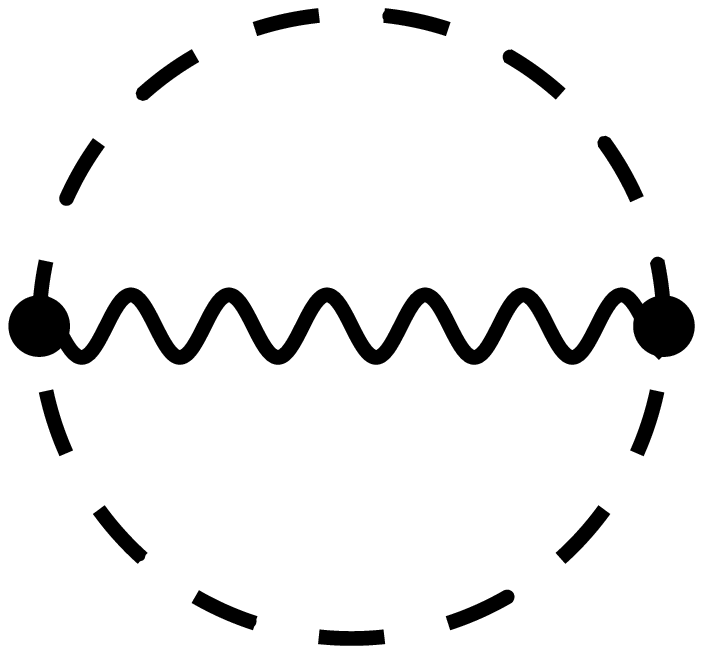}}
\put(4.7,5.8){\includegraphics[scale=0.2]{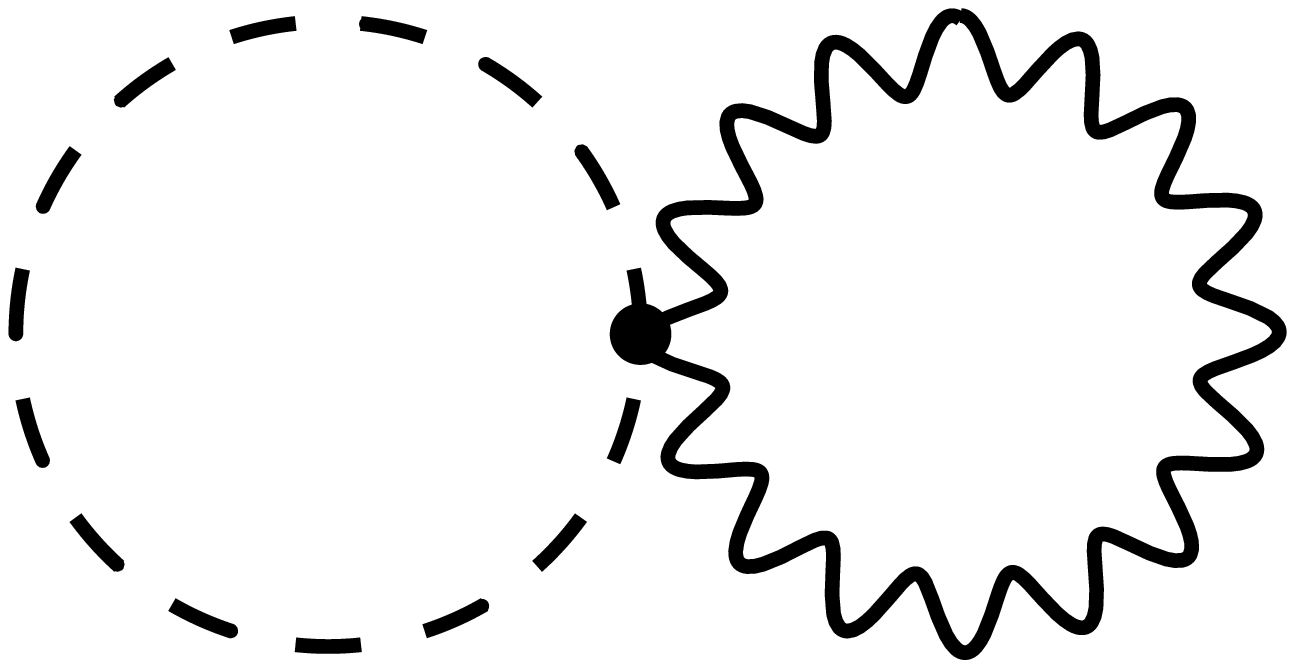}}
\put(9.8,5.8){\includegraphics[scale=0.2]{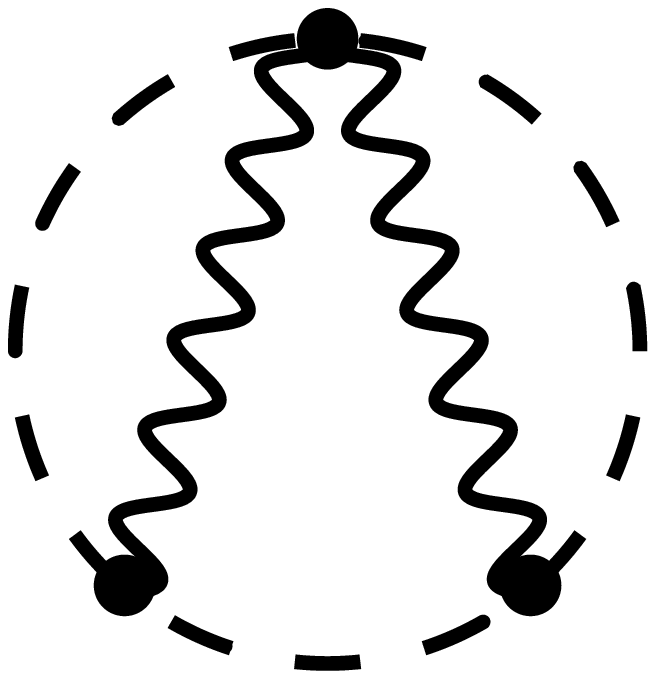}}
\put(13.6,5.8){\includegraphics[scale=0.2]{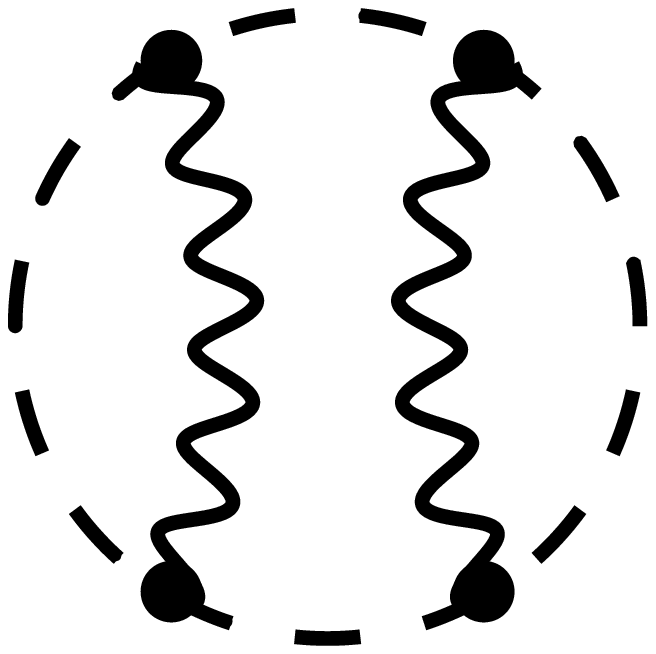}}

\put(1.2,2.9){\includegraphics[scale=0.2]{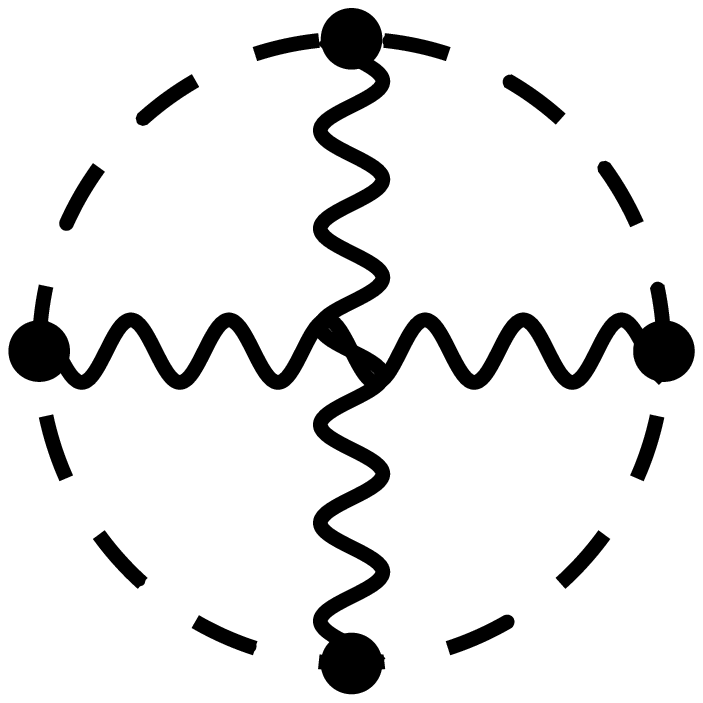}}
\put(4.3,2.9){\includegraphics[scale=0.2]{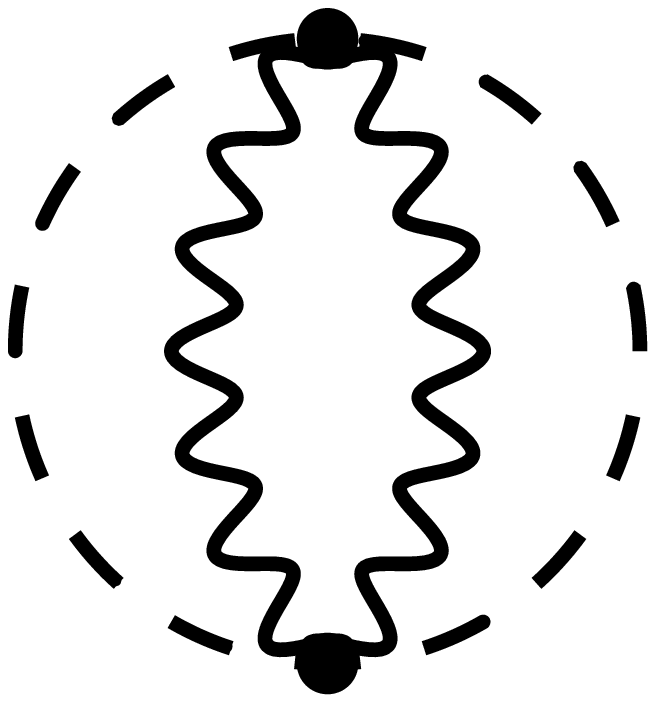}}
\put(7.4,2.85){\includegraphics[scale=0.2]{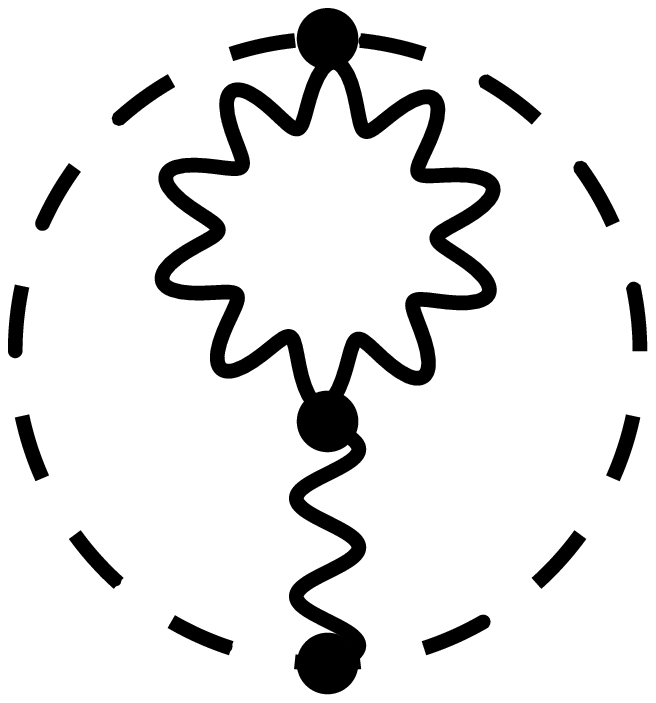}}
\put(10.5,2.9){\includegraphics[scale=0.2]{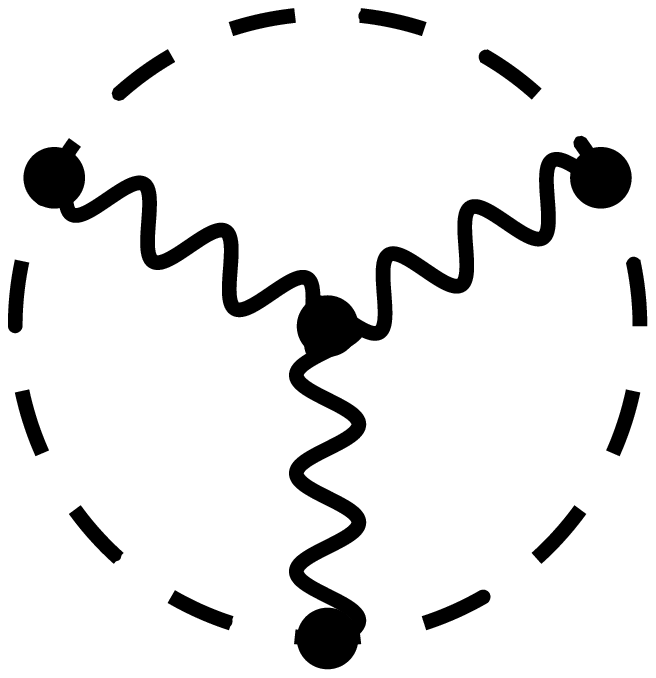}}
\put(13.6,2.9){\includegraphics[scale=0.20]{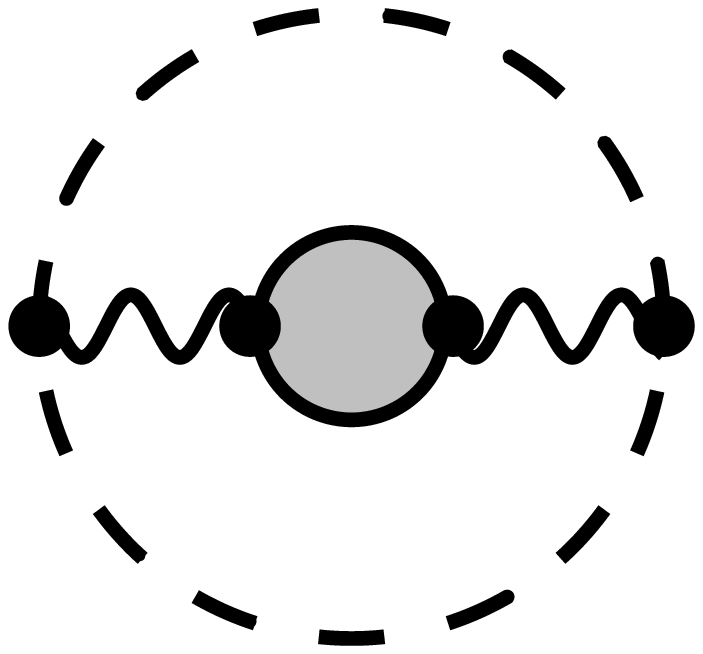}}

\put(1.1,0){\includegraphics[scale=0.19]{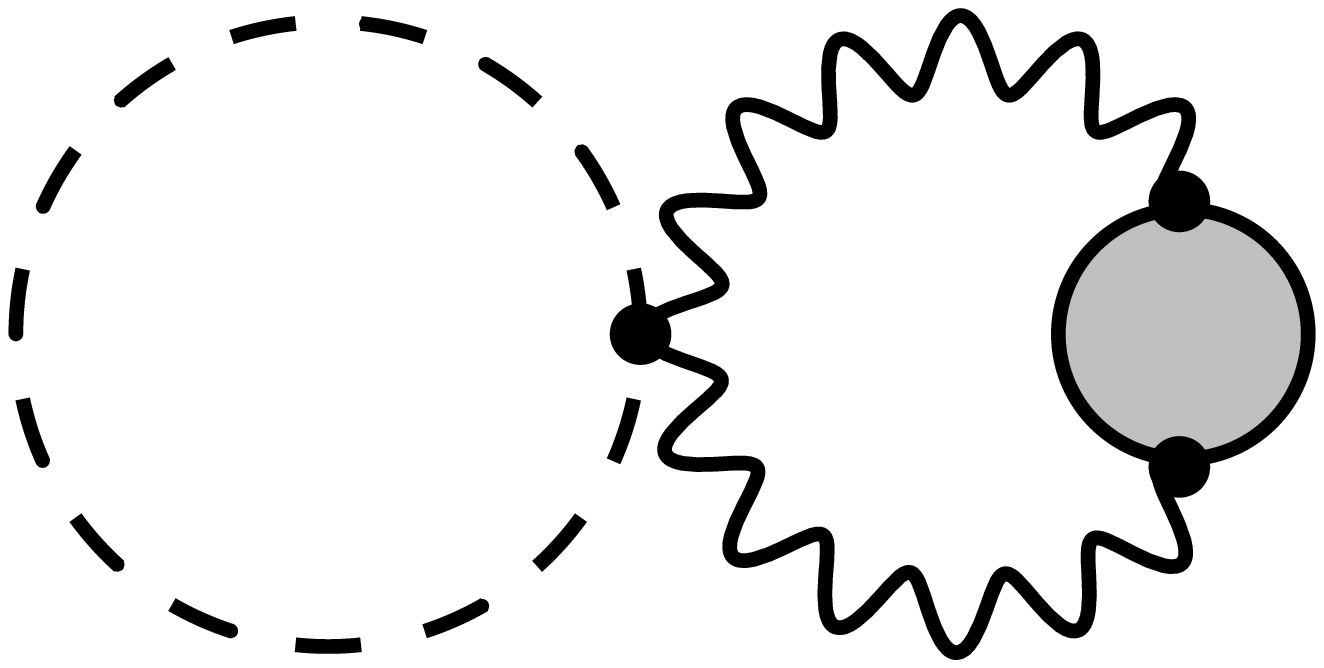}}
\put(4.9,0){\includegraphics[scale=0.19]{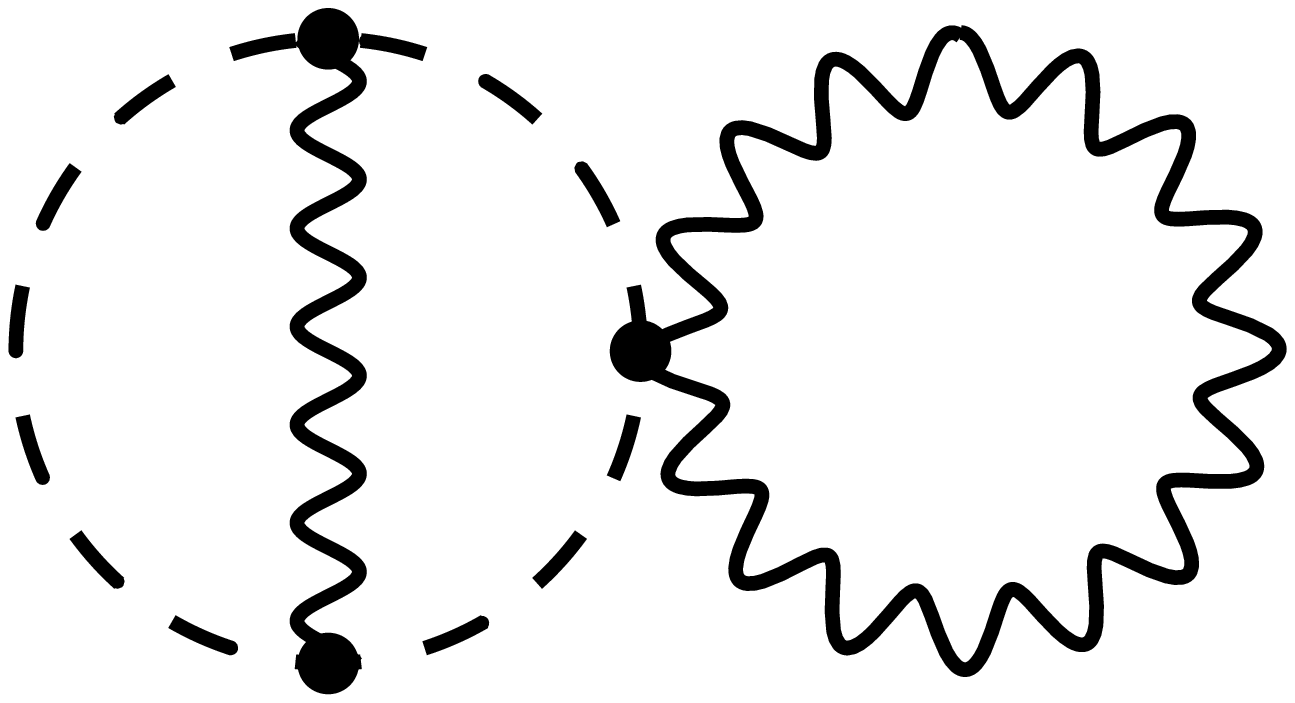}}
\put(8.5,0){\includegraphics[scale=0.23]{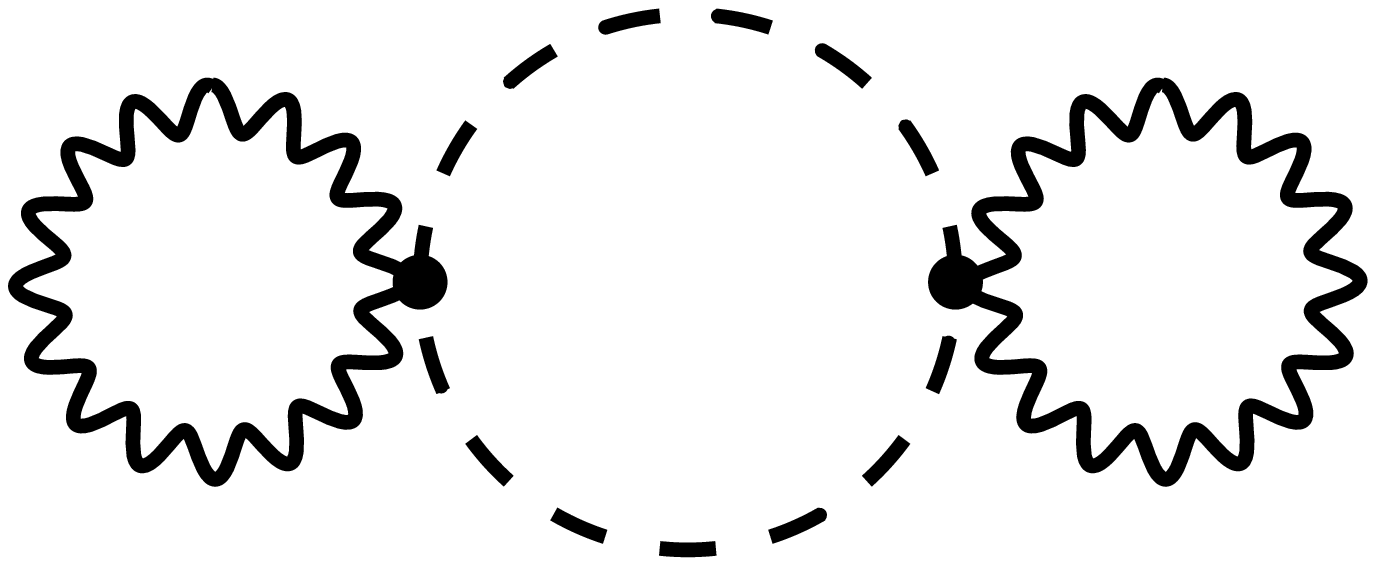}}
\put(13.0,0){\includegraphics[scale=0.2]{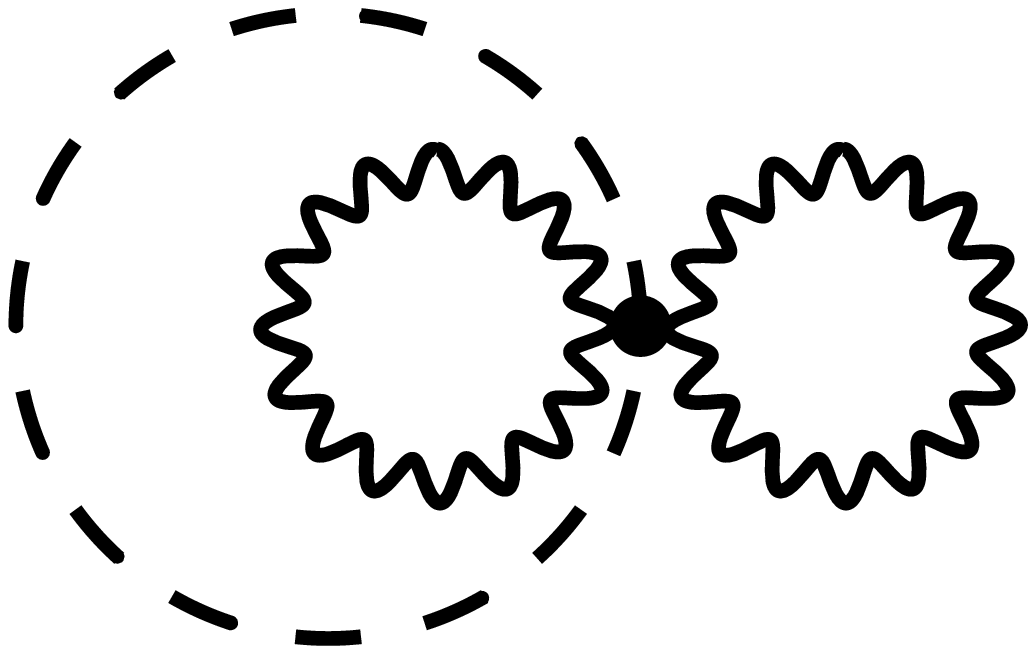}}

\put(0.8,7.3){B1}
\put(4.4,7.3){B2}
\put(9.3,7.3){B3}
\put(13.2,7.3){B4}

\put(0.8,4.4){B5}
\put(3.9,4.4){B6}
\put(7.0,4.4){B7}
\put(10.1,4.4){B8}
\put(13.2,4.4){B9}

\put(0.8,1.5){B10}
\put(4.5,1.5){B11}
\put(8.5,1.5){B12}
\put(12.6,1.5){B13}
\end{picture}
\caption{Supergraphs containing ghost loops which are essential for calculating the three-loop $\beta$-function. Note that the supergraphs containing two ghost loops should be multiplied by the factor $1/2$.}
\label{Figure_Ghost_Beta_Diagrams}
\end{figure}

\begin{figure}[h]
\begin{picture}(0,4)
\put(0.1,1){\includegraphics[scale=0.17]{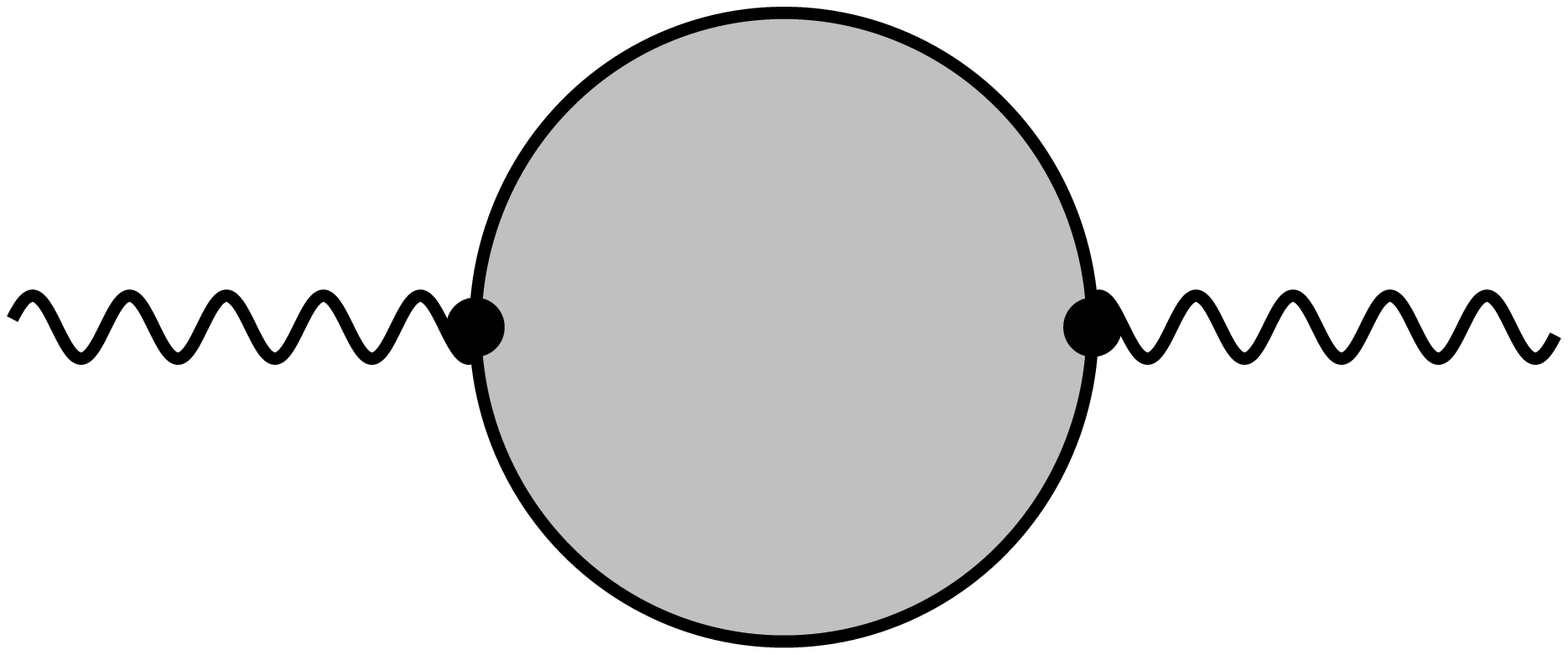}}
\put(3.7,1.6){$= \ \left\{\vphantom{\begin{array}{c} 1\\ 1\\ 1\\ 1\\ 1\\ 1\\ 1\\ 1 \end{array}} \hspace*{11cm}\right\}$}
\put(4.8,2.4){\includegraphics[scale=0.35]{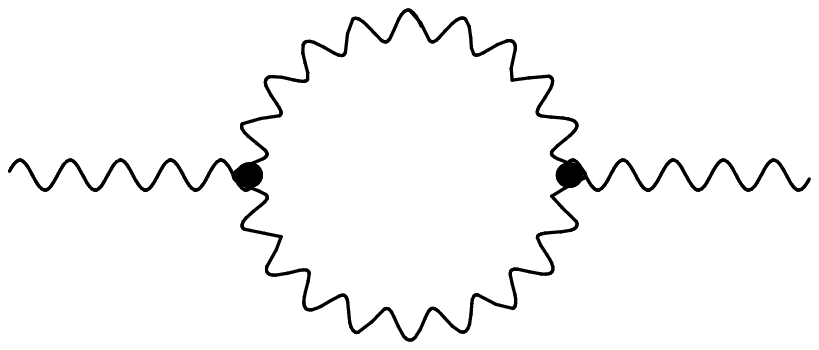}}
\put(5.2,0){\includegraphics[scale=0.35]{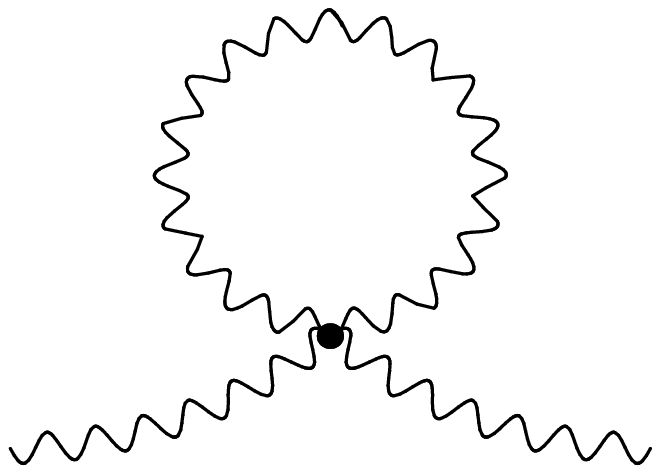}}
\put(8.7,2.43){\includegraphics[scale=0.35]{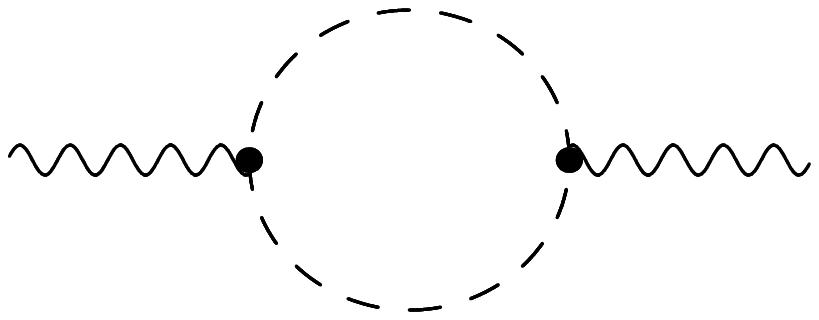}}
\put(9.0,-0.01){\includegraphics[scale=0.35]{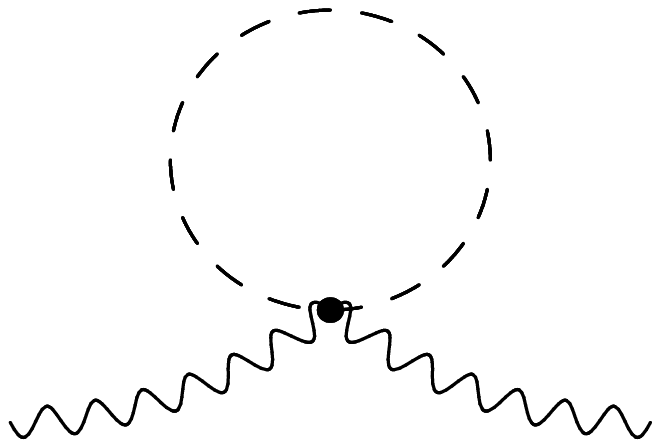}}
\put(12.6,2.4){\includegraphics[scale=0.35]{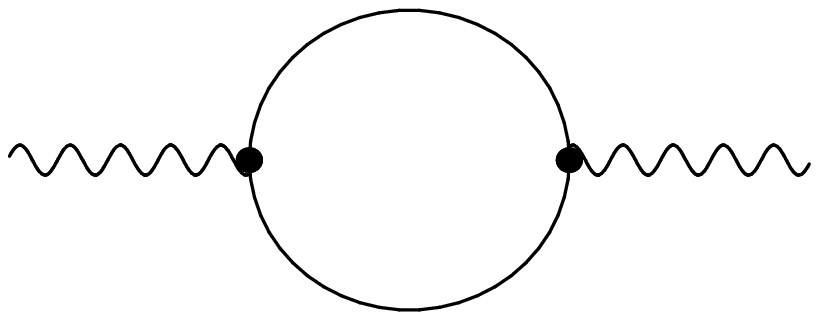}}
\put(13.0,-0.02){\includegraphics[scale=0.35]{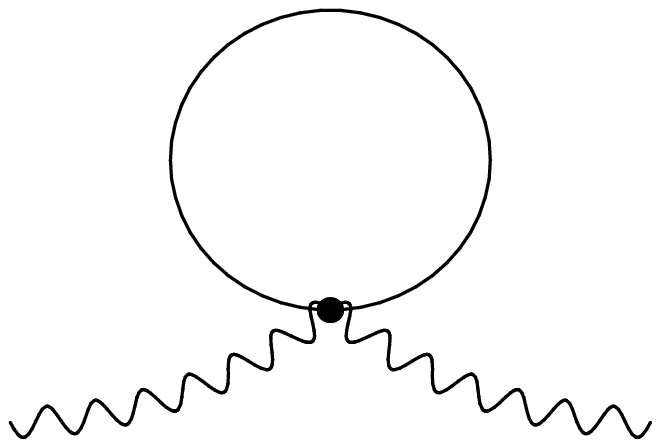}}
\put(5.0,3.6){V1} \put(8.9,3.6){V2} \put(12.8,3.6){V3}
\put(5.0,1.3){V4} \put(8.9,1.3){V5} \put(12.8,1.3){V6}
\end{picture}
\caption{Superdiagrams contributing to the one-loop polarization operator of the quantum gauge superfield.}
\label{Figure_Polarization_Operator}
\end{figure}

Now, let us apply the method described in the previous section for calculating the three-loop contributions to the $\beta$-function coming from the supergraphs containing loops of the Faddeev--Popov ghosts. They are presented in Fig.~\ref{Figure_Ghost_Beta_Diagrams}. The gray circle in the graphs B9 and B10 denote the insertion of the one-loop polarization operator. The corresponding superdiagrams (depicted in Fig. \ref{Figure_Polarization_Operator}) have been calculated in Ref. \cite{Kazantsev:2017fdc}. However, it is important that for the superdiagrams containing two ghost loops the effective supergraphs B9 and B10 produce an extra factor of 2. That is why such superdiagrams should be multiplied by the factor $1/2$. The loops of the Pauli--Villars superfields in Fig. \ref{Figure_Polarization_Operator} are denoted by the same solid line as the loops of the usual matter superfield $\phi_i$. Constructing the graphs in Fig. \ref{Figure_Ghost_Beta_Diagrams} we took into account that in the considered approximation the vertices with two ghost legs and three legs of the quantum gauge superfield are absent because

\begin{equation}
\frac{V}{1-e^{\pm 2V}} = \mp \frac{1}{2} + \frac{1}{2} V \mp \frac{1}{6} V^2 \pm \frac{1}{90} V^4 + O(V^6).
\end{equation}

\noindent
In higher loops they can possibly be essential due to the nonlinear renormalization of the quantum gauge superfield. However, in this paper calculating the $\beta$-function we take into account the parameters of the nonlinear renormalization only in the two-loop approximation. Namely, the vertex in the superdiagram B2 contains the term

\begin{equation}\label{Nonlinear_Vertex}
 -\frac{3}{4} e_0^2\, y_0\, G^{ABCD} \int d^4x\,d^4\theta\, (\bar c^A + \bar c^{*A}) V^C V^D (c^B-c^{*B}),
\end{equation}

\noindent
which is very important for calculating RGFs, see Ref. \cite{Kazantsev:2018kjx} for details.

\begin{figure}[h]
\begin{picture}(0,9.5)
\put(0.7,7.9){\includegraphics[scale=0.14]{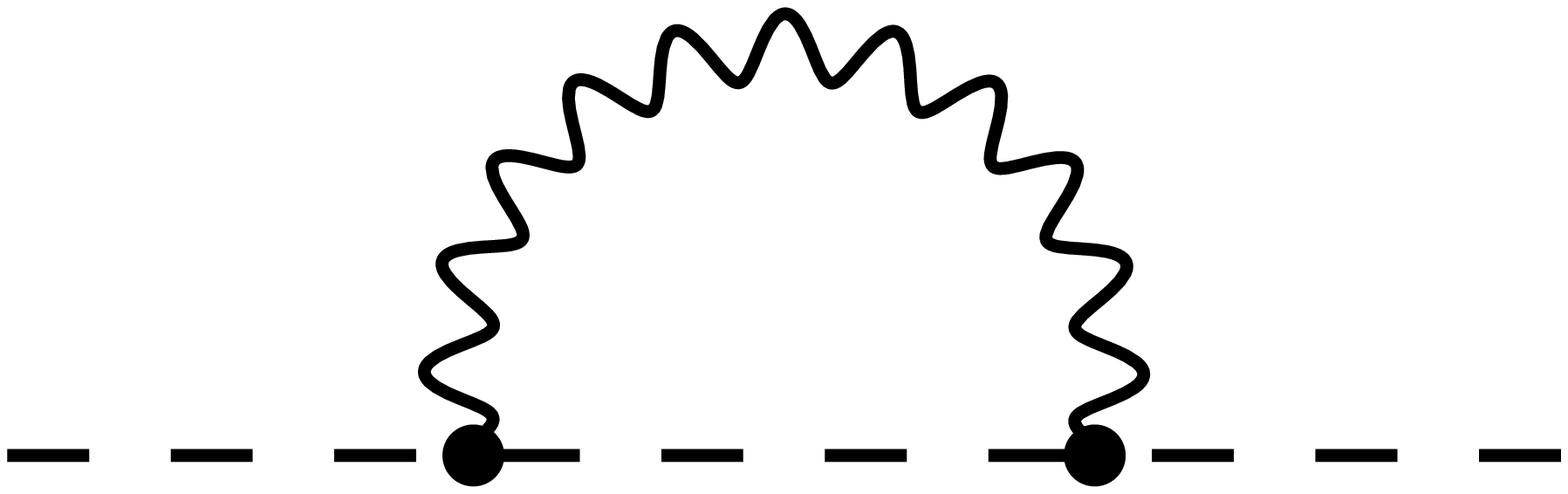}}
\put(0.7,9.0){$A1$}
\put(4.7,7.9){\includegraphics[scale=0.14]{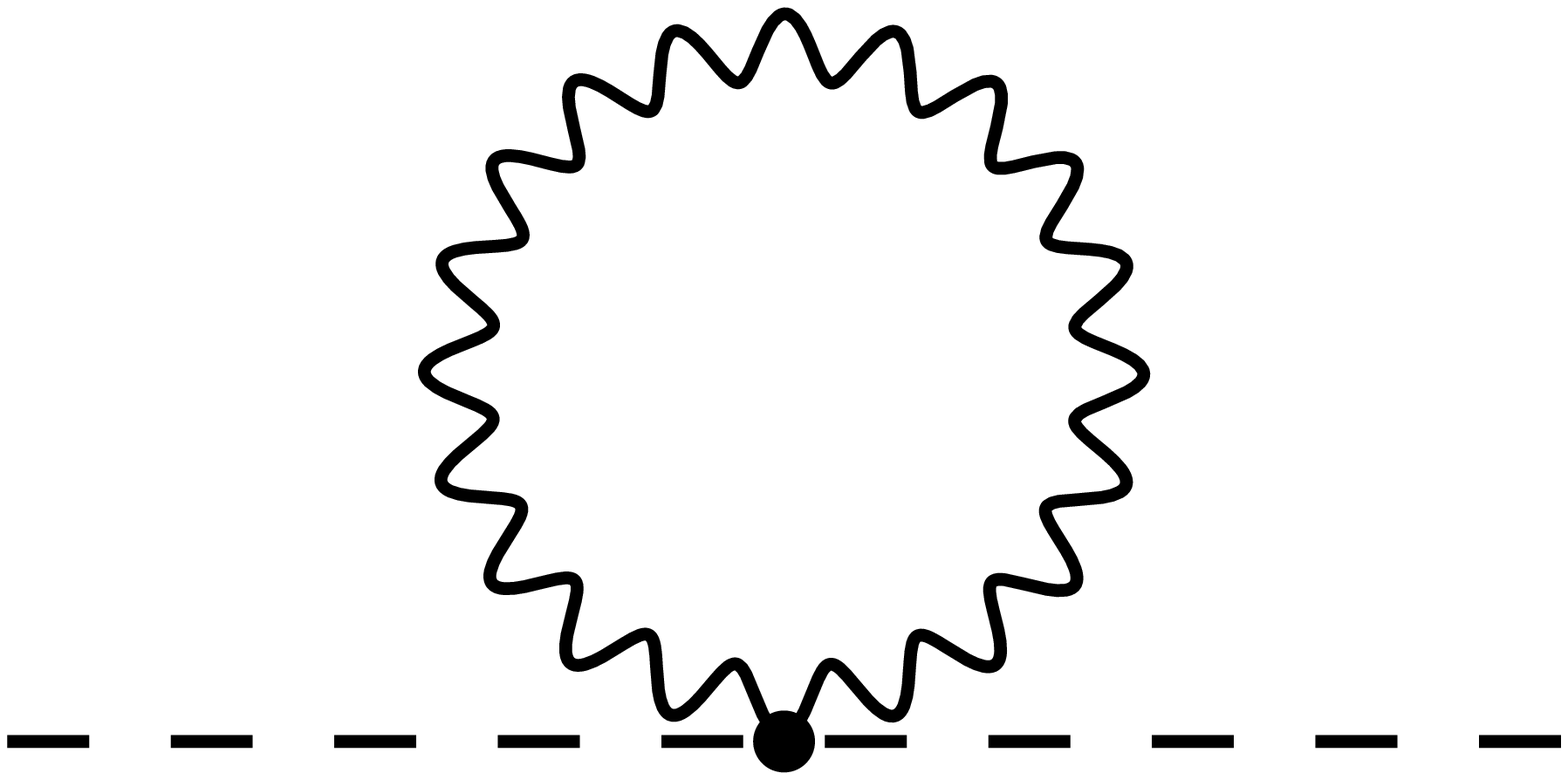}}
\put(4.7,9.0){$A2$}
\put(8.7,7.9){\includegraphics[scale=0.14]{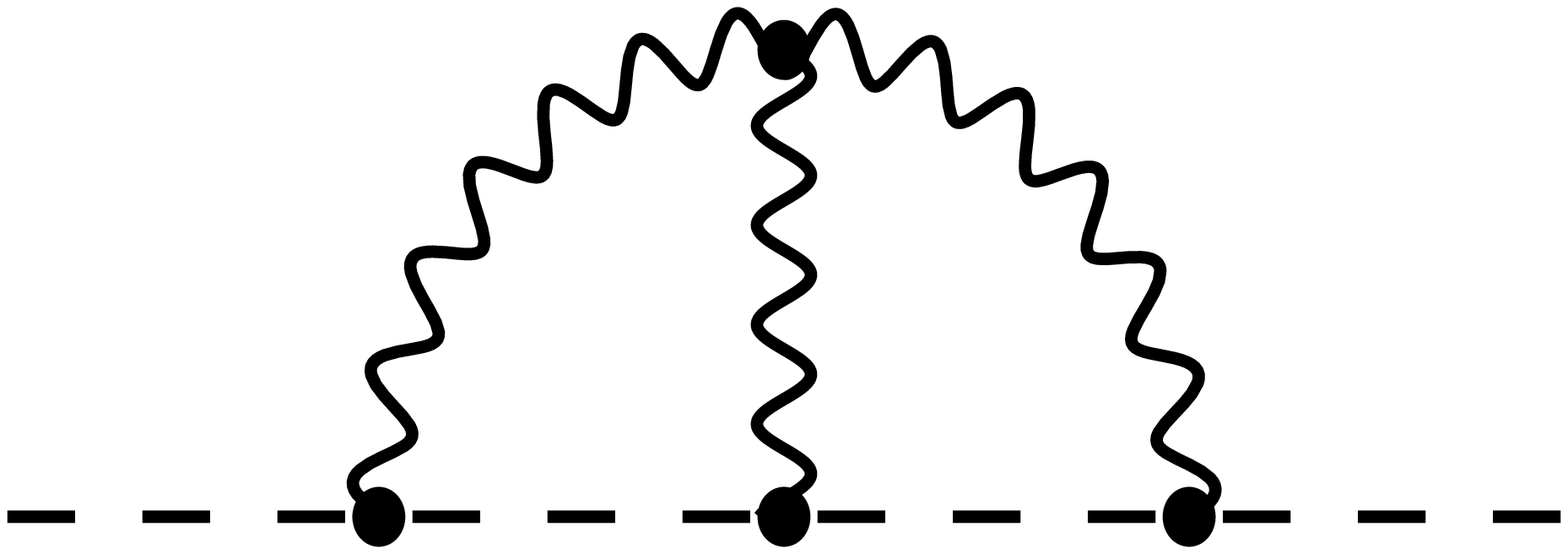}}
\put(8.7,9.0){$A3$}
\put(12.7,7.9){\includegraphics[scale=0.14]{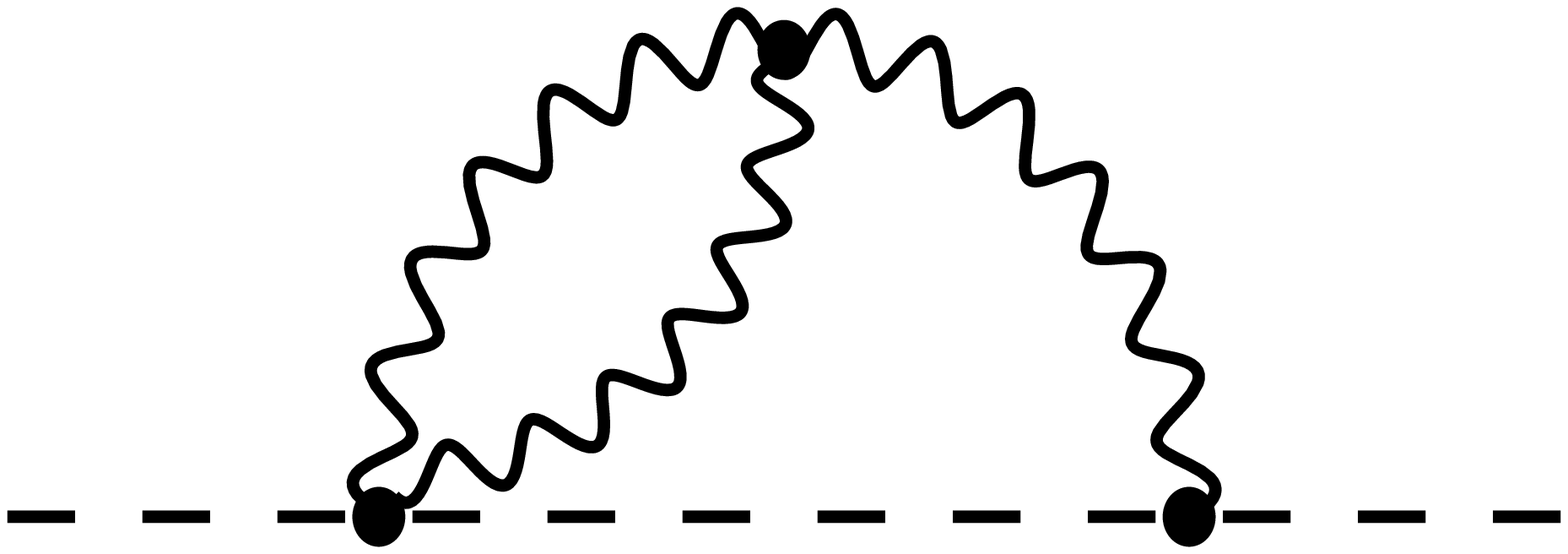}}
\put(12.7,9.0){$A4$}

\put(0.7,5.6){\includegraphics[scale=0.14]{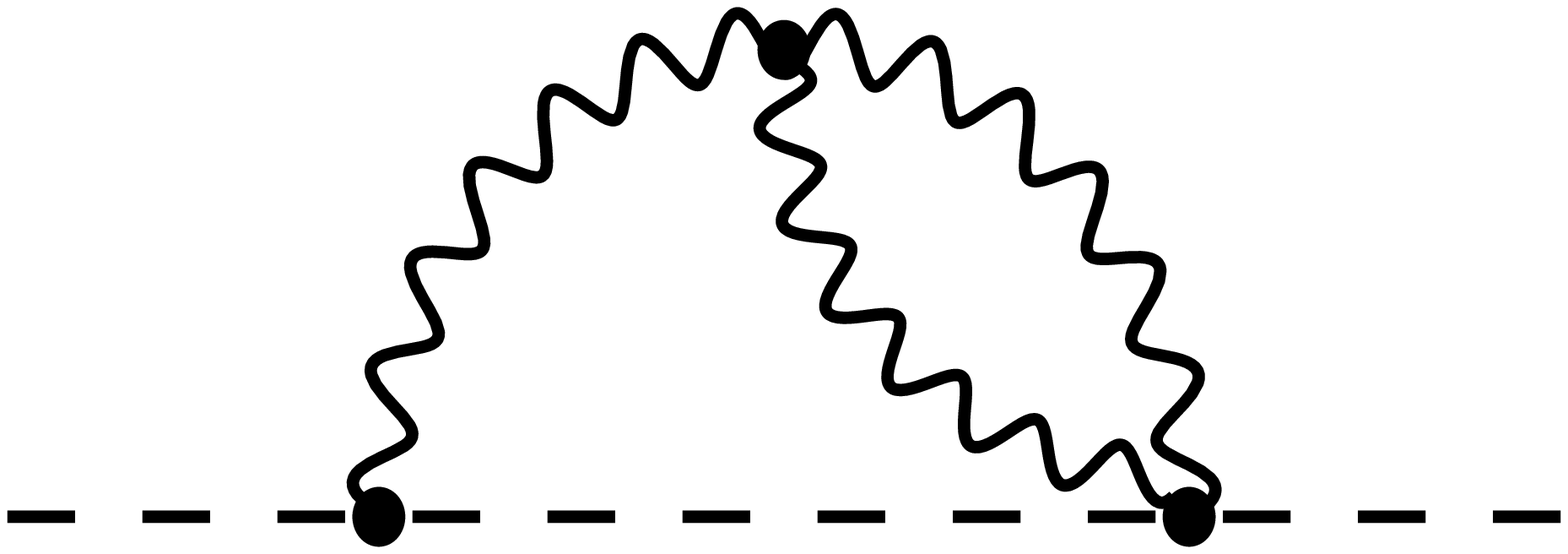}}
\put(0.7,6.5){$A5$}
\put(4.7,5.15){\includegraphics[scale=0.14]{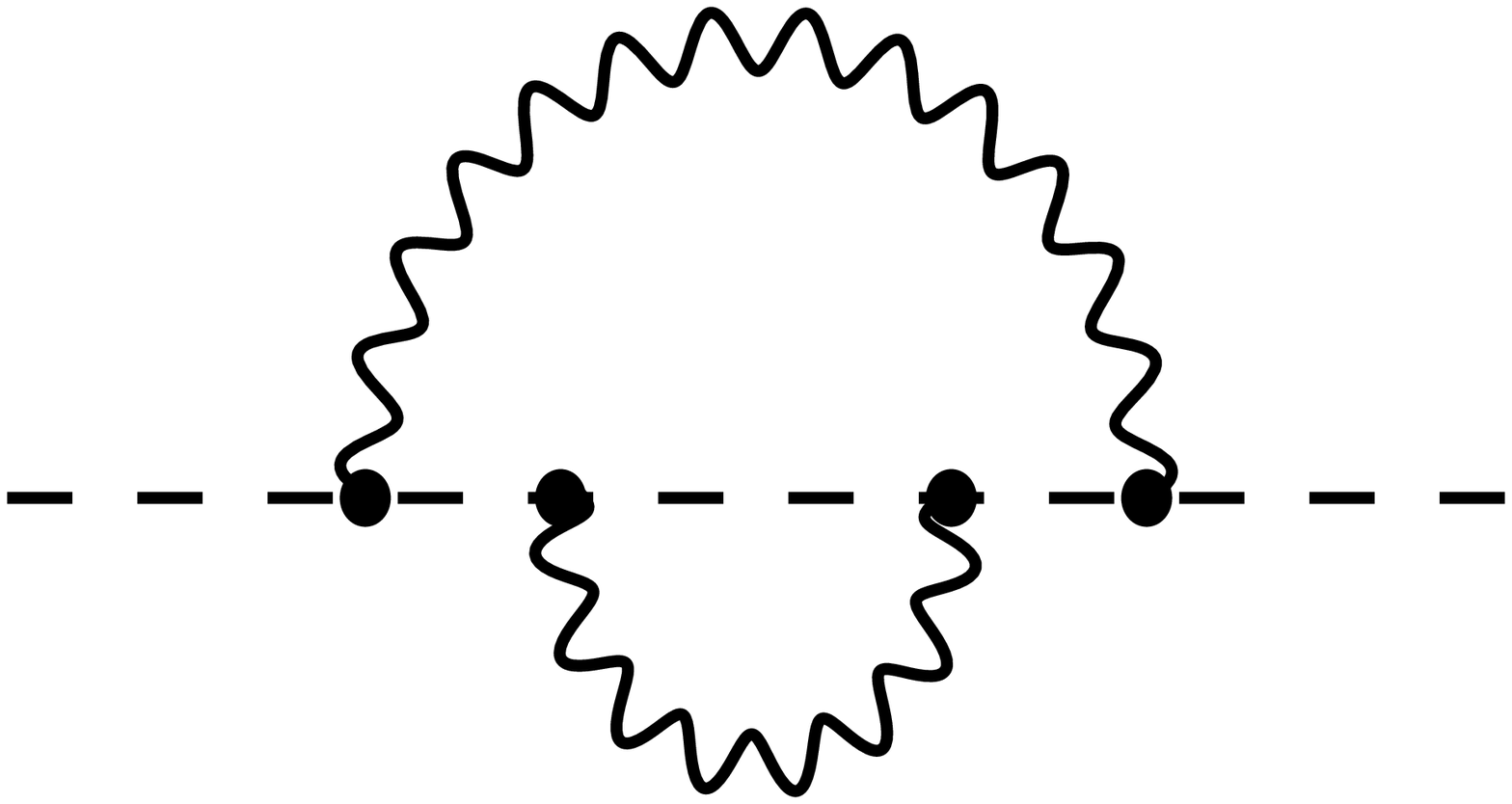}}
\put(4.7,6.5){$A6$}
\put(8.7,5.1){\includegraphics[scale=0.14]{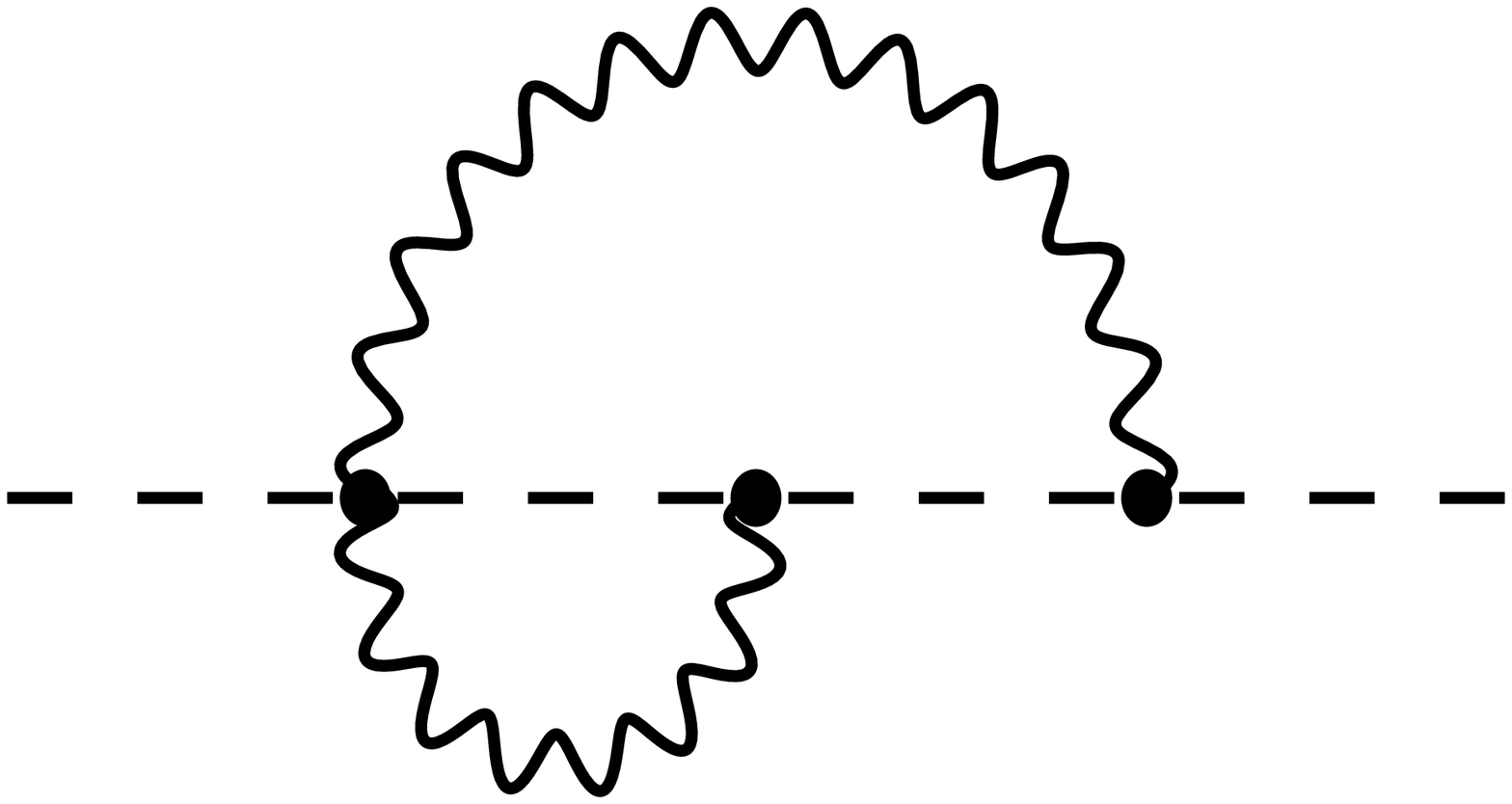}}
\put(8.7,6.5){$A7$}
\put(12.7,5.1){\includegraphics[scale=0.14]{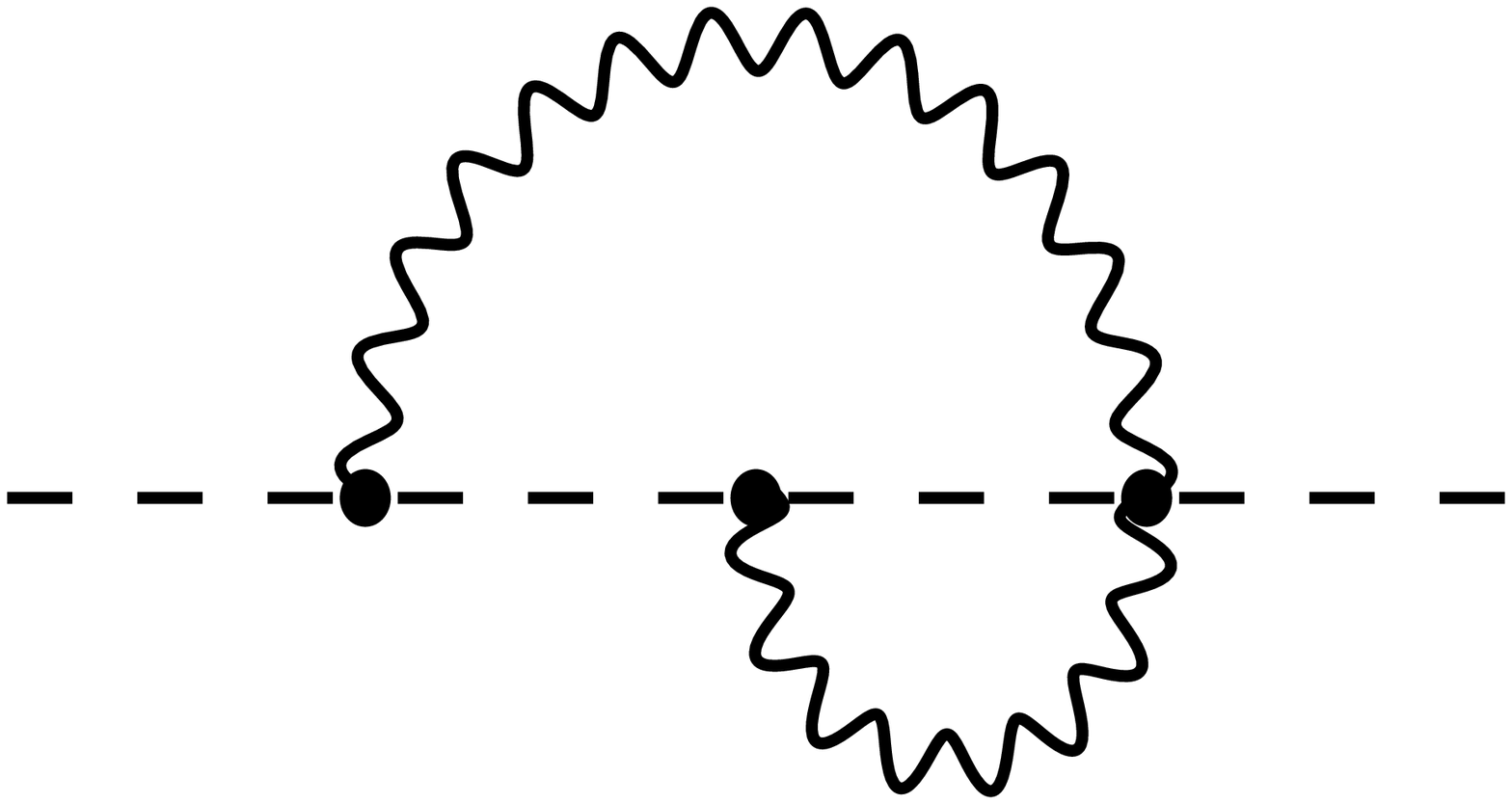}}
\put(12.7,6.5){$A8$}

\put(0.7,2.6){\includegraphics[scale=0.14]{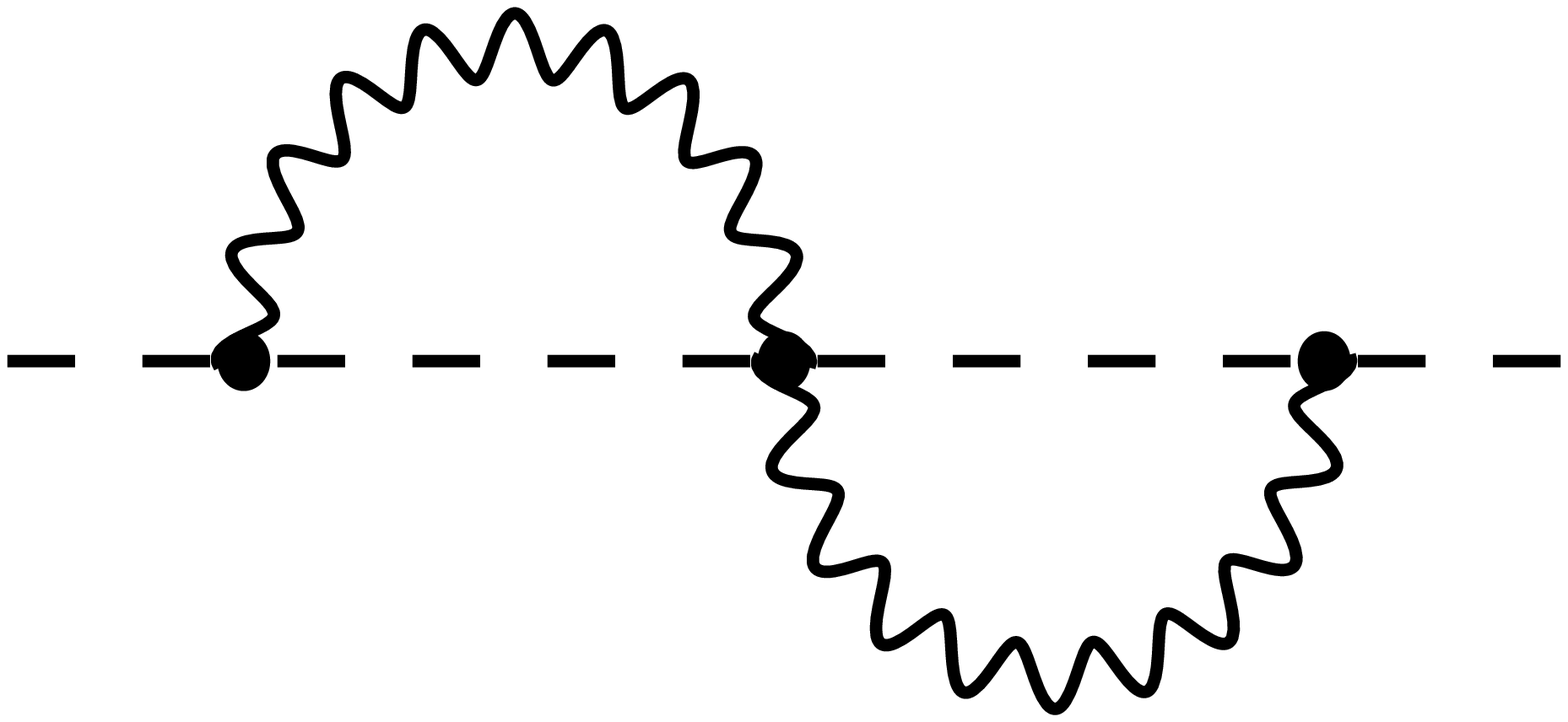}}
\put(0.7,4.0){$A9$}
\put(4.7,2.5){\includegraphics[scale=0.14]{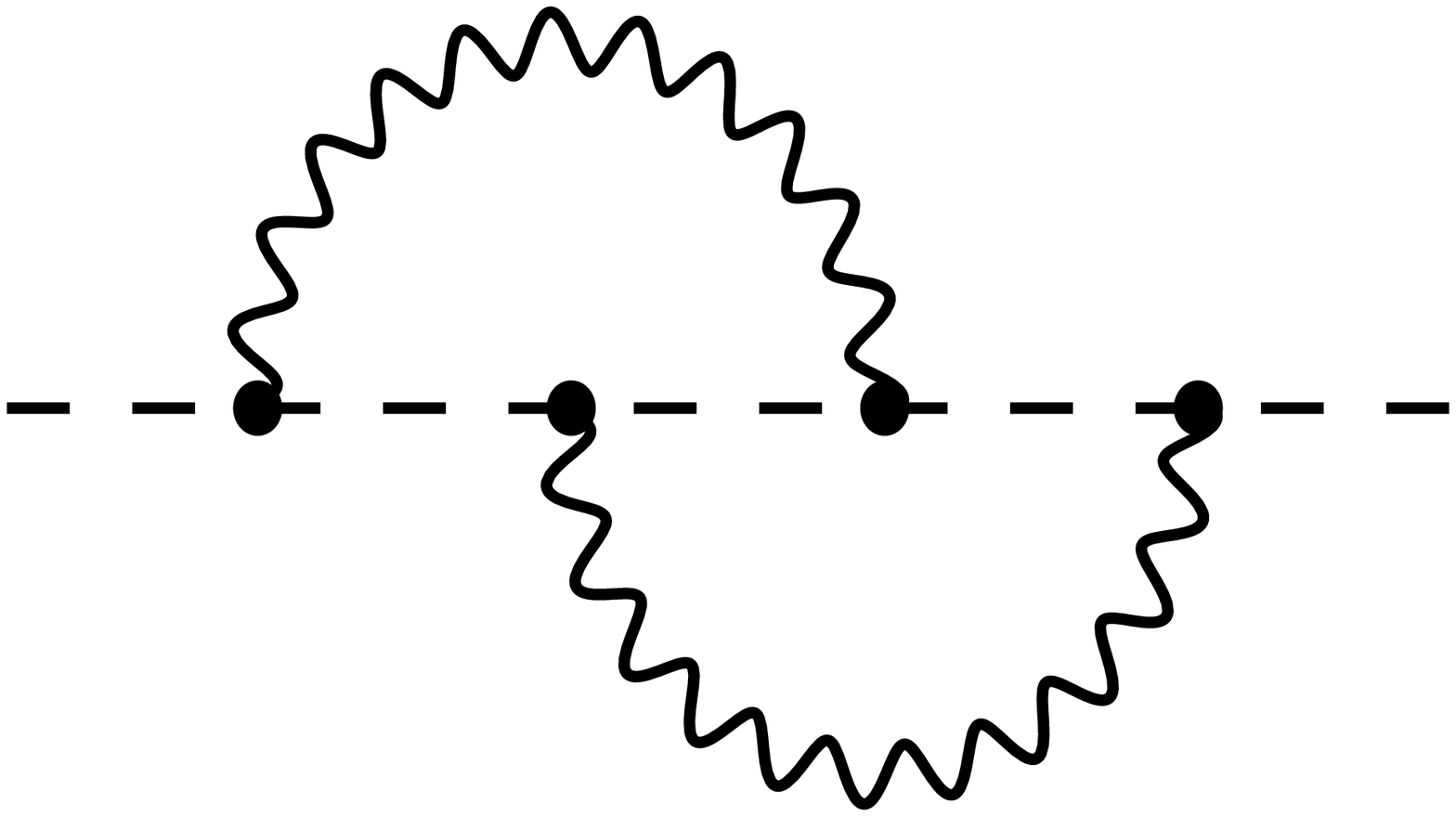}}
\put(4.7,4.0){$A10$}
\put(8.7,2.5){\includegraphics[scale=0.14]{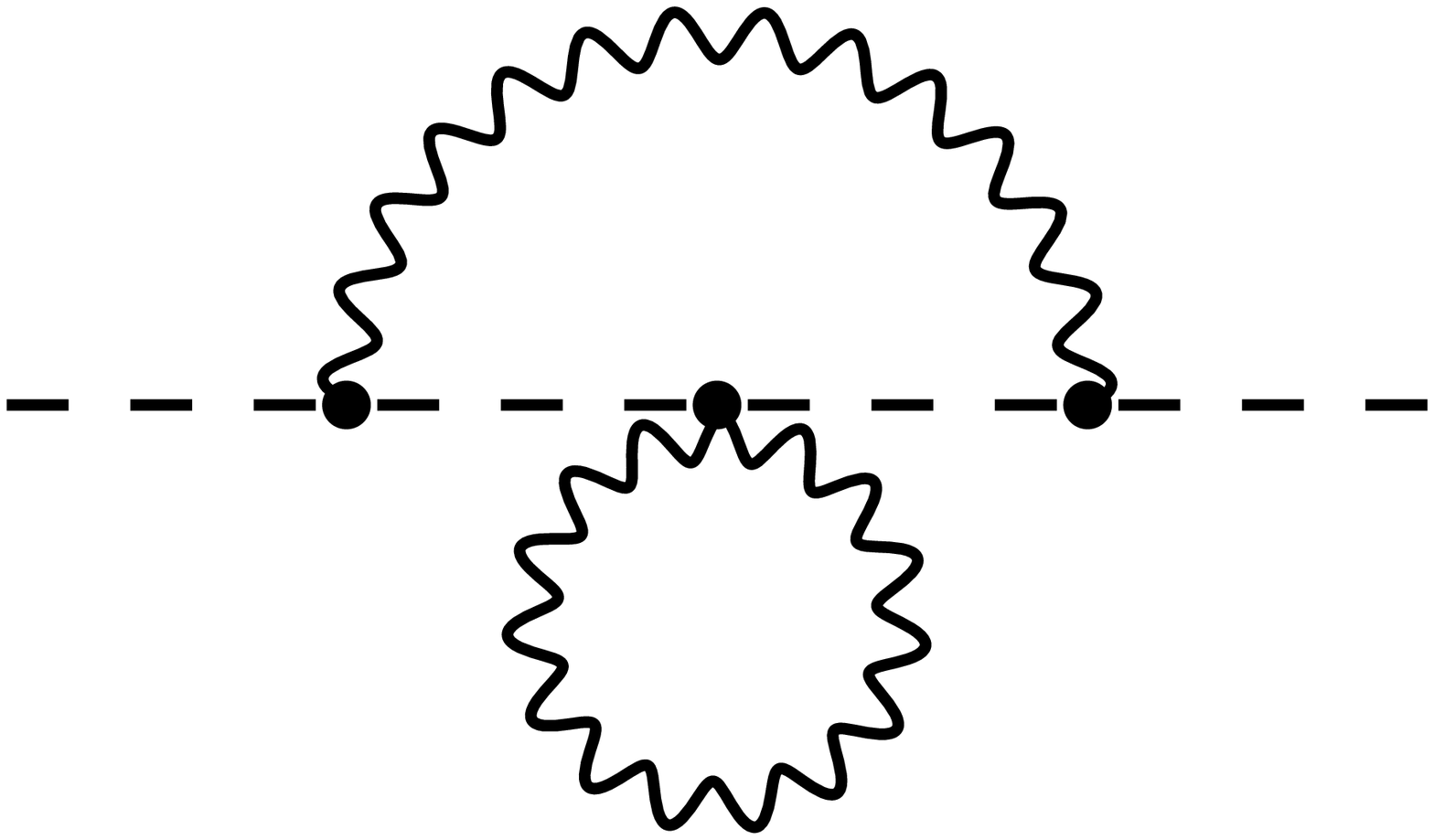}}
\put(8.7,4.1){$A11$}
\put(12.7,2.67){\includegraphics[scale=0.14]{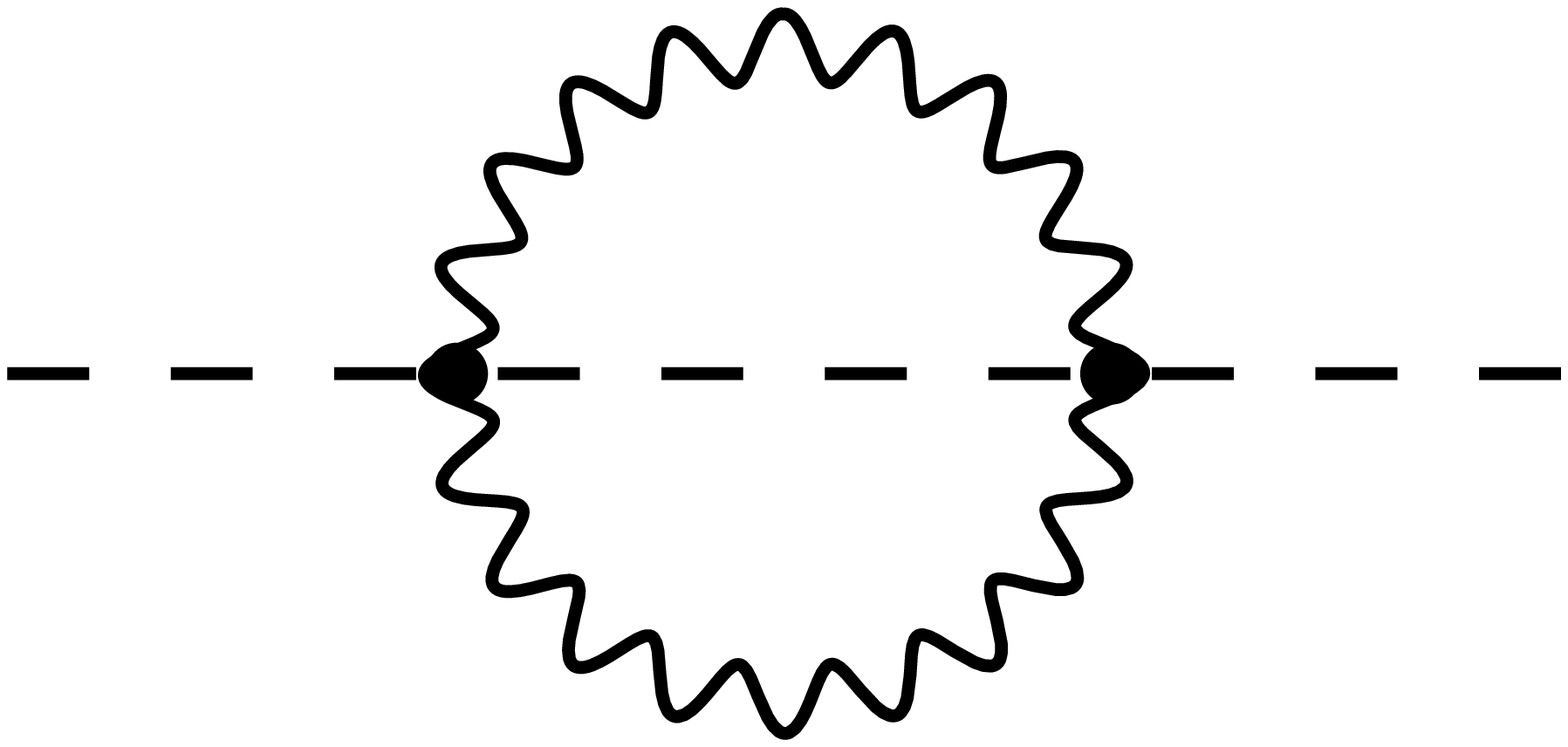}}
\put(12.7,4.1){$A12$}

\put(2.9,-0.3){\includegraphics[scale=0.14]{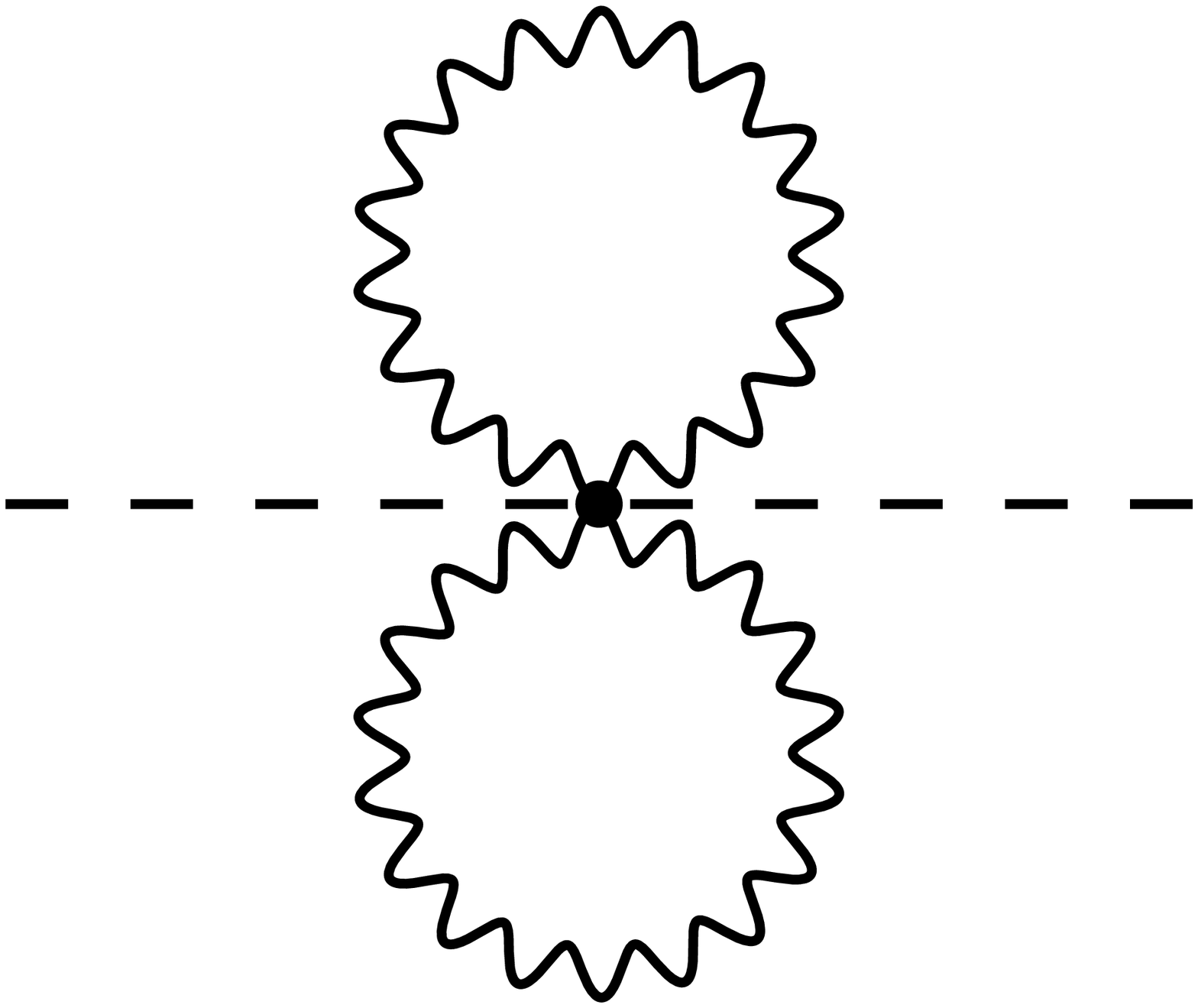}}
\put(2.9,1.6){$A13$}
\put(6.9,0.1){\includegraphics[scale=0.14]{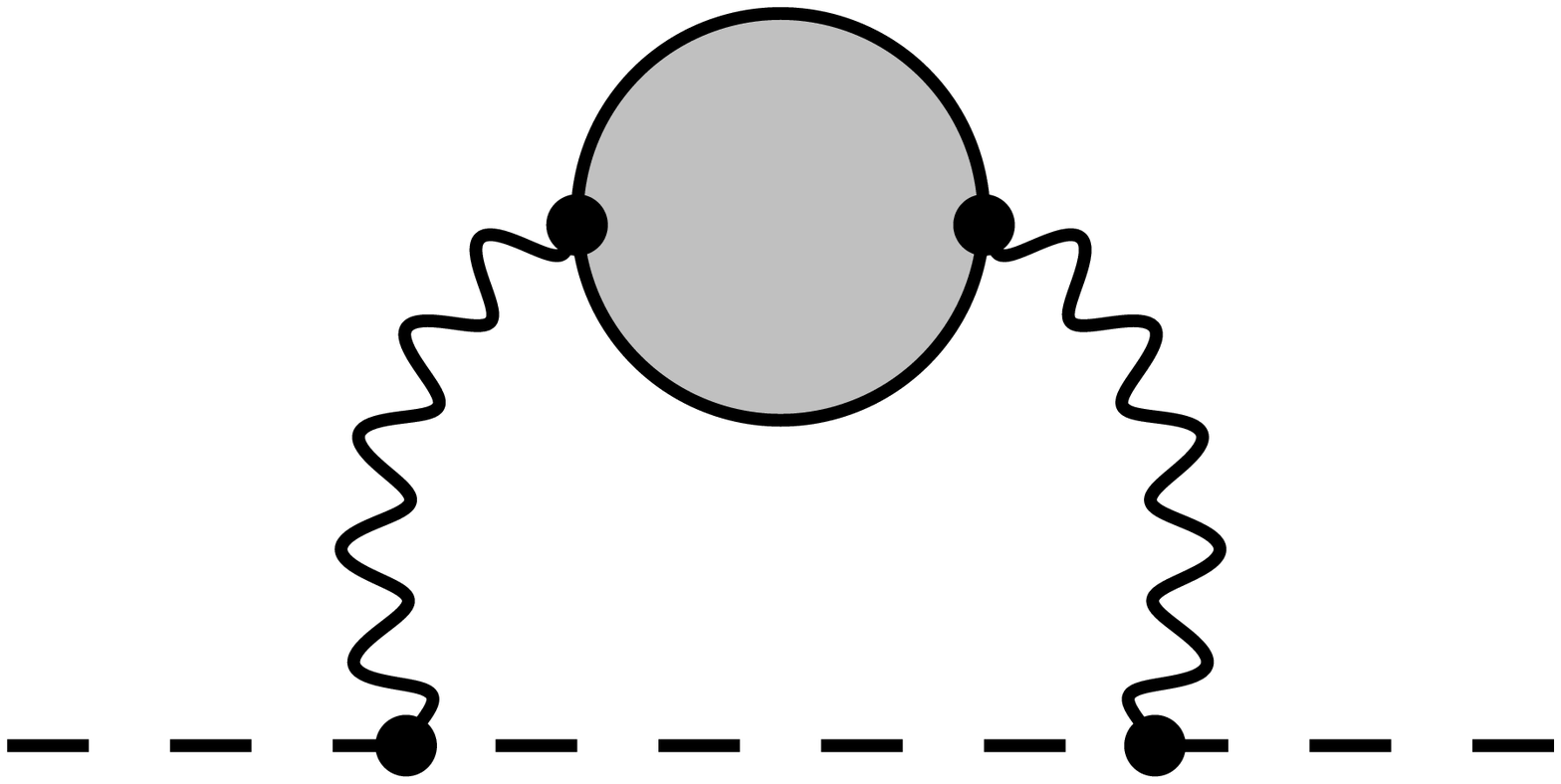}}
\put(6.9,1.6){$A14$}
\put(10.9,0.1){\includegraphics[scale=0.14]{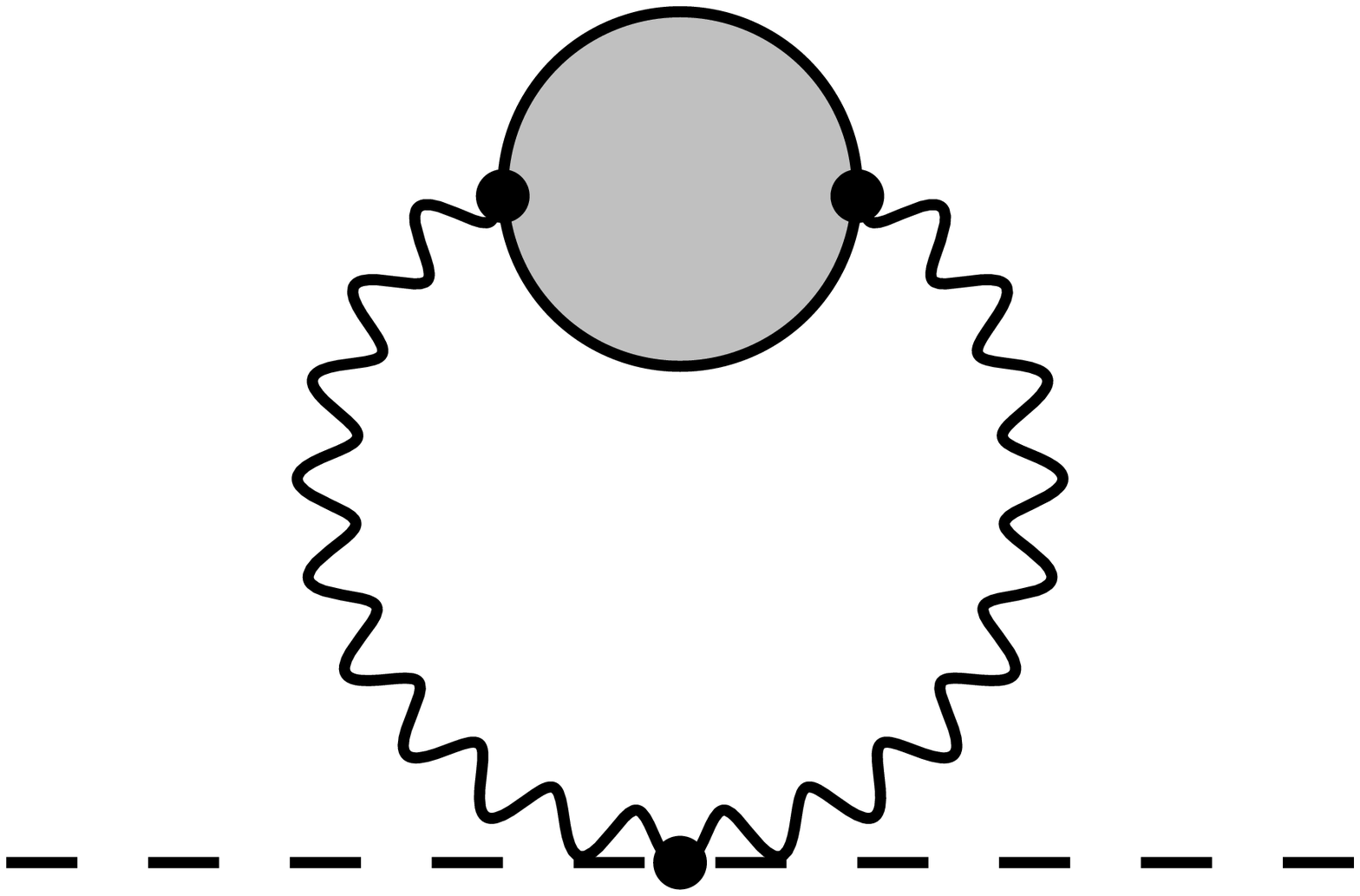}}
\put(10.9,1.6){$A15$}
\end{picture}
\caption{One- and two-loop superdiagrams contributing to the two-point Green function of the Faddeev--Popov ghosts.}\label{Figure_Ghost_Gamma_Diagrams}
\end{figure}

\begin{figure}[h]
\begin{picture}(0,3)
\put(2,0){\includegraphics[scale=0.18]{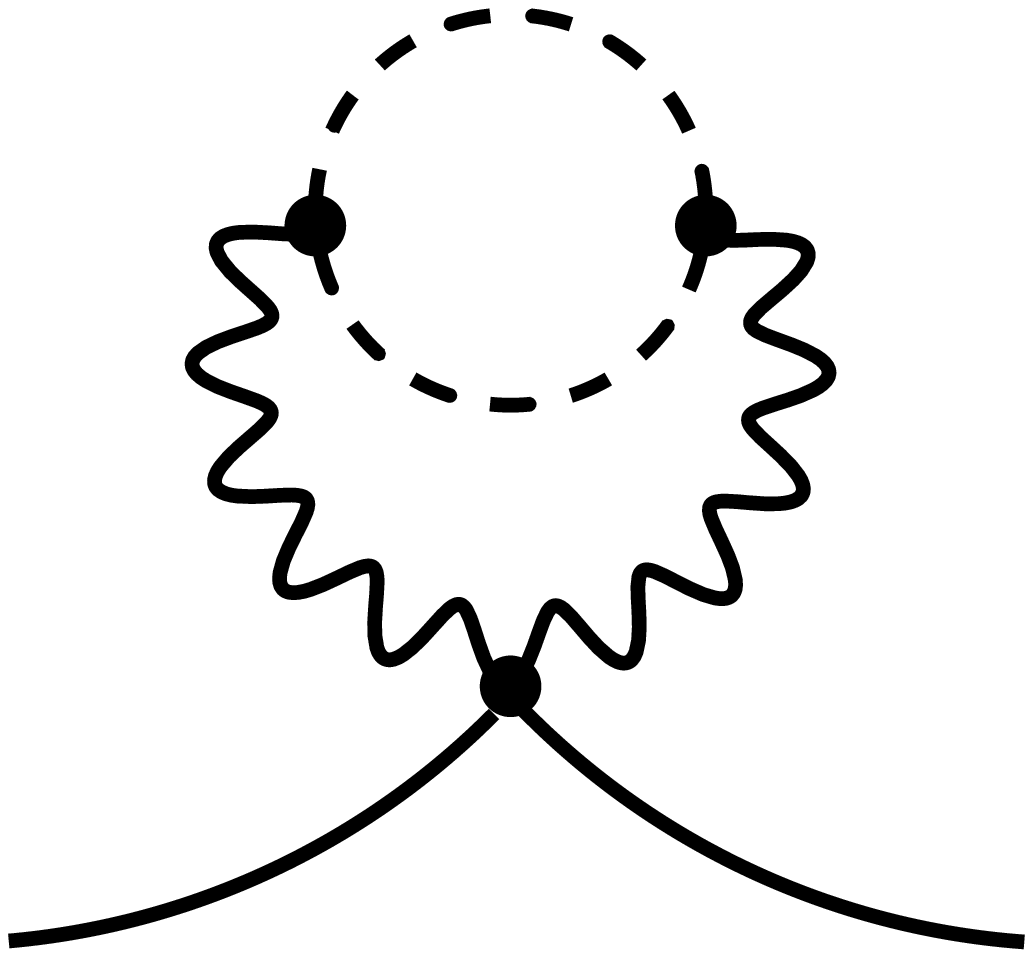}}
\put(5.1,0){\includegraphics[scale=0.18]{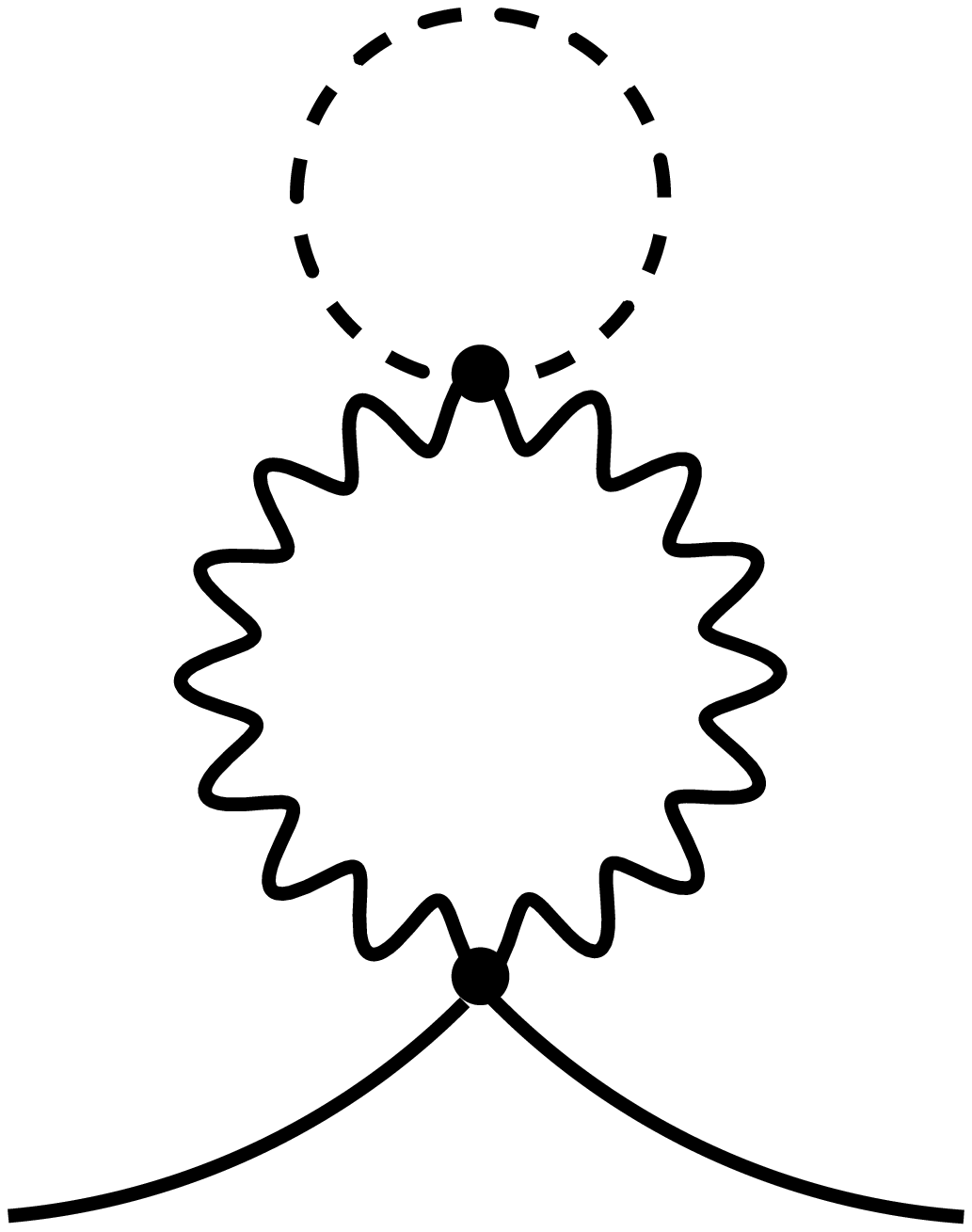}}
\put(8.2,0.2){\includegraphics[scale=0.18]{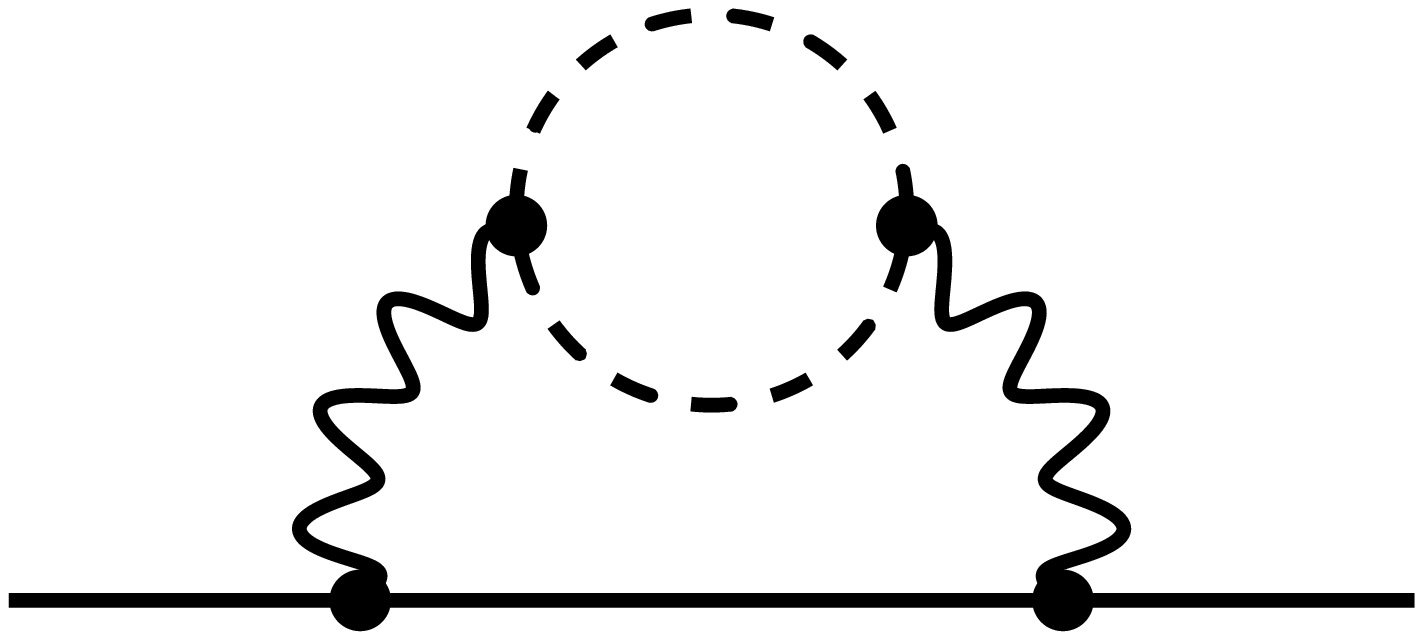}}
\put(11.8,0.2){\includegraphics[scale=0.18]{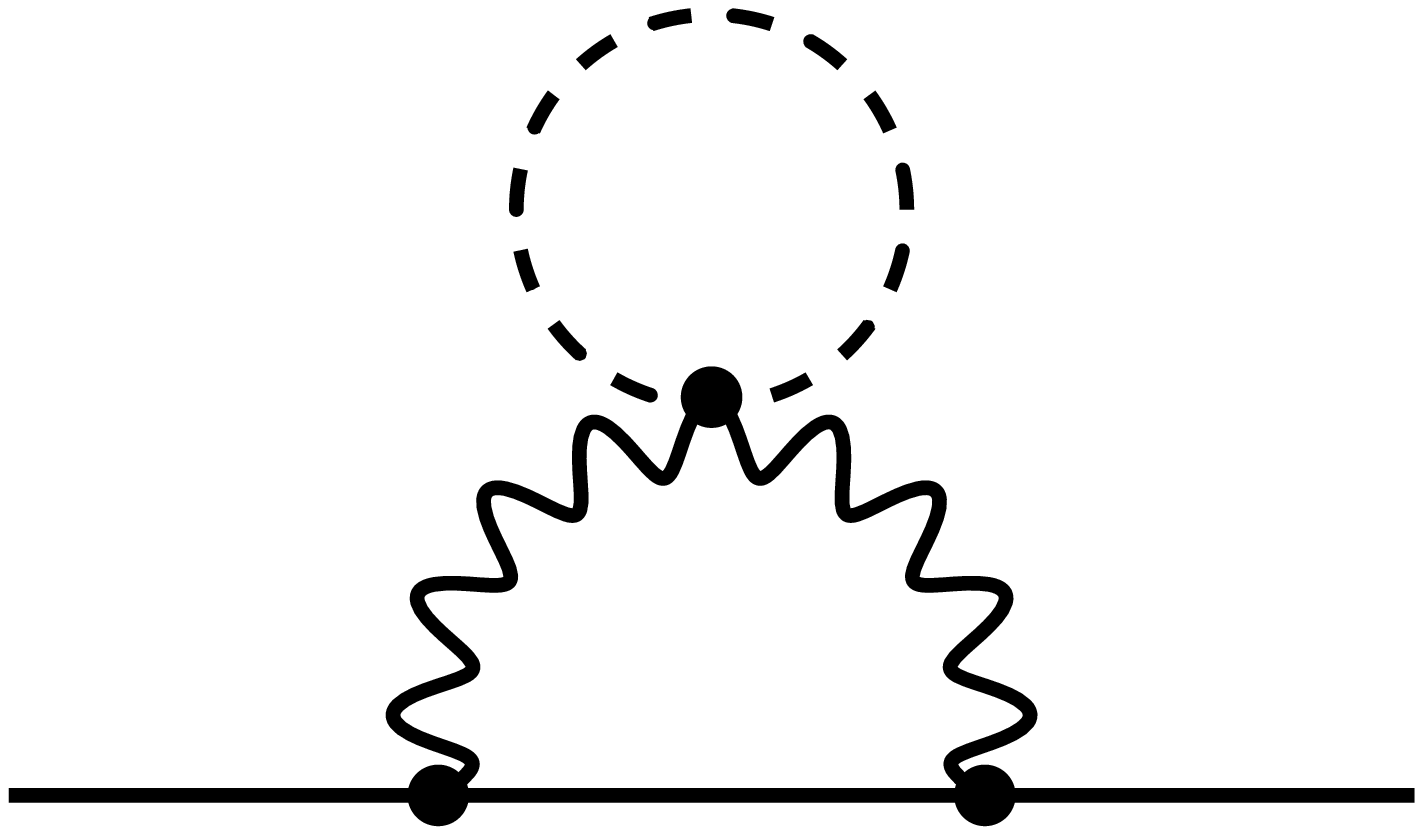}}
\put(1.8,2){M1} \put(4.9,2){M2} \put(8.5,2){M3} \put(11.9,2){M4}
\end{picture}
\caption{Two-loop superdiagrams with a ghost loop contributing to the anomalous dimension of the matter superfields.}\label{Figure_Matter_Anomalous_Dimension}
\end{figure}

As usual, to construct the superdiagrams contributing to the two-point function of the superfield $\bm{V}$, one should attach two external lines to the supergraphs presented in Fig. \ref{Figure_Ghost_Beta_Diagrams} in all possible ways. According to Ref. \cite{Stepanyantz:2016gtk}, their contributions to the $\beta$-function (defined in terms of the bare couplings) are possibly related to superdiagrams obtained by cutting internal lines in the original supergraphs by equations analogous to Eq. (\ref{NSVZ_Equation_New_Form}). In the considered case we will obtain one- and two-loop superdiagrams with two external ghost lines presented in Fig.~\ref{Figure_Ghost_Gamma_Diagrams} and the superdiagrams containing a ghost loop contributing to the anomalous dimensions of the matter superfields (see Fig.~\ref{Figure_Matter_Anomalous_Dimension}) and of the quantum gauge superfield.

The superdiagrams contributing to the anomalous dimension of the quantum gauge superfield will not be considered in this paper because of the following reason: Due to the Slavnov--Taylor identities the two-point Green function of the quantum gauge superfield is transversal,

\begin{equation}
\Gamma^{(2)}_V - S^{(2)}_{\mbox{\scriptsize gf}} = -\frac{1}{2e_0^2} \mbox{tr}\int \frac{d^4p}{(2\pi)^4} d^4\theta\, V(-p,\theta)\, \partial^2\Pi_{1/2} V(p,\theta)\, G_V(\alpha_0,\lambda_0,Y_0,\Lambda/p).
\end{equation}

\noindent
However, contributions of various separate superdiagrams are not transversal,

\begin{eqnarray}\label{Separate_Graph}
&& \Delta\Gamma^{(2)}_V = -\frac{1}{2e_0^2} \mbox{tr}\int \frac{d^4p}{(2\pi)^4} d^4\theta\, \Big(V(-p,\theta)\, \partial^2\Pi_{1/2} V(p,\theta)\, \Delta G_V(\alpha_0,\lambda_0,Y_0,\Lambda/p)\qquad\nonumber\\
&& + V(-p,\theta)\, V(p,\theta)\, \Delta \widetilde G_V(\alpha_0,\lambda_0,Y_0,\Lambda/p)\Big).
\end{eqnarray}

\noindent
To obtain the transversal result, one should find the sum of all graphs. Then all longitudinal contributions $\Delta \widetilde G_V$ cancel each other, while the sum of the transversal parts will be related to the anomalous dimension $\gamma_V$. However, cutting internal gauge lines in a certain vacuum supergraph we obtain structures related to both $\Delta G_V$ and $\Delta\widetilde G_V$ in the resulting two-point superdiagrams. If we would like to compare the result with $\gamma_V$, then it would be necessary either to extract a part corresponding to $\Delta G_V$ or to deal with the sum of superdiagrams in which all longitudinal terms cancel each other. In this paper we consider only supergraphs containing ghost loops, but do not consider purely gauge supergraphs. However, the part of $\Gamma^{(2)}_V$ corresponding to the sum of all superdiagrams with ghost loops is not transversal. The transversal result is obtained only after adding the purely gauge contribution. That is why the part of the $\beta$-function coming from the supergraphs presented in Fig.~\ref{Figure_Ghost_Beta_Diagrams}
cannot be directly compared with the corresponding contributions to $\gamma_V$ produced by cutting internal gauge lines.

However, one can compare various parts of the $\beta$-function with the corresponding parts of $\gamma_c$ and $(\gamma_\phi)_j{}^i$. For this purpose we construct the expression for the contribution to the function $\beta/\alpha_0^2$ in the form of an integral of double total derivatives according to the algorithm described in Sect. \ref{Section_Algorithm}. The internal lines are cut due to the identity

\begin{equation}\label{Delta_Identity}
\frac{\partial^2}{\partial Q_\mu^2}\Big(\frac{1}{Q^2}\Big) = -4\pi^2 \delta^4(Q).
\end{equation}

\noindent
The cuts of the internal ghost and matter lines produce the terms containing $\gamma_c$ and $(\gamma_\phi)_j{}^i$, respectively, while the cuts of the gauge superfield propagators (corresponding to $\gamma_V$) are not taken into consideration in the present paper. Using this procedure we can match the supergraphs presented in Fig.~\ref{Figure_Ghost_Beta_Diagrams} and the ones in Figs. \ref{Figure_Ghost_Gamma_Diagrams} and \ref{Figure_Matter_Anomalous_Dimension},

\begin{eqnarray}\label{Cuttings}
&& \mbox{B1}\ \to\ \mbox{A1} + \ldots;\qquad\qquad\qquad\qquad\qquad\ \, \mbox{B8}\ \to\ \mbox{A3} + \ldots; \vphantom{\frac{1}{2}}\nonumber\\
&& \mbox{B2}\ \to\ \mbox{A2} + \ldots; \qquad\qquad\qquad\qquad\qquad\ \, \mbox{B9}\ \to\ \mbox{A14} + \mbox{M1} + \mbox{M3} + \ldots;\qquad \vphantom{\frac{1}{2}}\nonumber\\
&& \mbox{B3}\ \to\ \mbox{A7} + \mbox{A8} + \mbox{A9} + \ldots; \qquad\qquad\quad\ \, \mbox{B10}\ \to\ \mbox{A15} + \mbox{M2} + \mbox{M4} + \ldots; \vphantom{\frac{1}{2}}\nonumber\\
&& \mbox{B4}\ \to\ (\mbox{A1})\times (\mbox{A1}) + \mbox{A6} + \ldots; \qquad\qquad \mbox{B11}\ \to\ (\mbox{A1})\times (\mbox{A2}) + \mbox{A11} + \ldots; \qquad\vphantom{\frac{1}{2}}\nonumber\\
&& \mbox{B5}\ \to\ \mbox{A10} + \ldots; \qquad\qquad\qquad\qquad\quad\ \ \ \mbox{B12}\ \to\ (\mbox{A2}) \times (\mbox{A2}) + \ldots; \vphantom{\frac{1}{2}}\nonumber\\
&& \mbox{B6}\ \to\ \mbox{A12} + \ldots; \qquad\qquad\qquad\qquad\quad\ \ \ \mbox{B13}\ \to\ \mbox{A13} + \ldots;\vphantom{\frac{1}{2}}\nonumber\\
&& \mbox{B7}\ \to\ \mbox{A4} + \mbox{A5}+\ldots \vphantom{\frac{1}{2}}
\end{eqnarray}

\noindent
Here dots denote the superdiagrams contributing to the anomalous dimension $\gamma_V$, which are not considered in this paper. The crosses in the equations corresponding to the diagrams B4, B11, and B12 mean that some cuts of internal lines produce diagrams which are not 1PI. In this case one should do more cuts, which give a certain number $k\ge 2$ of 1PI graphs. They correspond to the terms with $k\ge 2$ in the expansion

\begin{equation}
\ln G = \ln (1+\Delta G) = \sum\limits_{k=1}^\infty \frac{(-1)^{k+1}}{k} (\Delta G)^k,
\end{equation}

\noindent
where $G$ is either $G_c$ or $(G_\phi)_j{}^i$ (in general, $G_V$ is also possible) defined by the equations

\begin{eqnarray}
&& \Gamma^{(2)}_c = \frac{1}{4}\int \frac{d^4q}{(2\pi)^4} d^4\theta\,\Big(c^{*A}(-q,\theta) \bar c^A(q,\theta) + \bar c^{*A}(-q,\theta) c^A(q,\theta)\Big) G_c(\alpha_0,\lambda_0,Y_0,\Lambda/q);\qquad\nonumber\\
&& \Gamma^{(2)}_\phi = \frac{1}{4}\int \frac{d^4q}{(2\pi)^4} d^4\theta\,\phi^{*j}(-q,\theta)\phi_i(q,\theta) \big(G_\phi\big)_j{}^i(\alpha_0,\lambda_0,Y_0,\Lambda/q).
\end{eqnarray}

The results for all supergraphs depicted in Fig.~\ref{Figure_Ghost_Beta_Diagrams} obtained by the method described in Sect.~\ref{Section_Algorithm} are collected in Appendix \ref{Appendix_Ghost_Supergraphs}. Namely, we present their contributions to the function $\beta/\alpha_0^2$ written in the form of integrals of double total derivatives, and the corresponding parts of $\gamma_c$. The main result of the calculation can be written in the form of the equations analogous to (\ref{Cuttings}),\footnote{For the supergraphs B9 and B10 we calculate only the sum, see the explanation in Appendix \ref{Appendix_Ghost_Supergraphs}.}

\begin{eqnarray}\label{Relation_First}
&&\hspace*{-7mm} \Delta_{\mbox{\scriptsize B1}}\Big(\frac{\beta}{\alpha_0^2}\Big) = \frac{C_2}{\pi} \frac{d}{d\ln\Lambda} \Big(\Delta_{\mbox{\scriptsize A1}} G_c + \ldots\Big)\Big|_{Q=0};\\
&&\hspace*{-7mm} \Delta_{\mbox{\scriptsize B2}}\Big(\frac{\beta}{\alpha_0^2}\Big) = \frac{C_2}{\pi} \frac{d}{d\ln\Lambda} \Big(\Delta_{\mbox{\scriptsize A2}} G_c + \ldots\Big)\Big|_{Q=0};\\
&&\hspace*{-7mm} \Delta_{\mbox{\scriptsize B3}}\Big(\frac{\beta}{\alpha_0^2}\Big) = \frac{C_2}{\pi} \frac{d}{d\ln\Lambda}\Big(\Delta_{\mbox{\scriptsize A7}} G_c + \Delta_{\mbox{\scriptsize A8}} G_c + \Delta_{\mbox{\scriptsize A9}} G_c + \ldots\Big)\Big|_{Q=0};\\
&&\hspace*{-7mm} \Delta_{\mbox{\scriptsize B4}}\Big(\frac{\beta}{\alpha_0^2}\Big) = \frac{C_2}{\pi} \frac{d}{d\ln\Lambda}\Big(-\frac{1}{2}\left(\Delta_{\mbox{\scriptsize A1}} G_c\right)^2 + \Delta_{\mbox{\scriptsize A6}} G_c + \ldots\Big)\Big|_{Q=0};\\
&&\hspace*{-7mm} \Delta_{\mbox{\scriptsize B5}}\Big(\frac{\beta}{\alpha_0^2}\Big) = \frac{C_2}{\pi} \frac{d}{d\ln\Lambda} \Big(\Delta_{\mbox{\scriptsize A10}} G_c + \ldots\Big)\Big|_{Q=0};\\
&&\hspace*{-7mm} \Delta_{\mbox{\scriptsize B6}}\Big(\frac{\beta}{\alpha_0^2}\Big) = \frac{C_2}{\pi} \frac{d}{d\ln\Lambda} \Big(\Delta_{\mbox{\scriptsize A12}} G_c + \ldots\Big)\Big|_{Q=0};\\
&&\hspace*{-7mm} \Delta_{\mbox{\scriptsize B7}}\Big(\frac{\beta}{\alpha_0^2}\Big) = \frac{C_2}{\pi} \frac{d}{d\ln\Lambda} \Big(\Delta_{\mbox{\scriptsize A4}} G_c + \Delta_{\mbox{\scriptsize A5}} G_c + \ldots\Big)\Big|_{Q=0};\\
&&\hspace*{-7mm} \Delta_{\mbox{\scriptsize B8}}\Big(\frac{\beta}{\alpha_0^2}\Big) = \frac{C_2}{\pi} \frac{d}{d\ln\Lambda} \Big(\Delta_{\mbox{\scriptsize A3}} G_c + \ldots\Big)\Big|_{Q=0};\\
\label{Relation_B910}
&&\hspace*{-7mm} \Delta_{\mbox{\scriptsize B9}}\Big(\frac{\beta}{\alpha_0^2}\Big) + \Delta_{\mbox{\scriptsize B10}}\Big(\frac{\beta}{\alpha_0^2}\Big)
= \frac{C_2}{\pi} \frac{d}{d\ln\Lambda} \Big(\Delta_{\mbox{\scriptsize A14}} G_c +\Delta_{\mbox{\scriptsize A15}} G_c +\ldots\Big) \nonumber\\
&&\hspace*{-7mm}\qquad -\frac{1}{2\pi r} C(R)_i{}^j \frac{d}{d\ln\Lambda} \Big(\big(\Delta_{\mbox{\scriptsize M1}} G_\phi\big)_j{}^i + \big(\Delta_{\mbox{\scriptsize M2}} G_\phi\big)_j{}^i
+ \big(\Delta_{\mbox{\scriptsize M3}} G_\phi\big)_j{}^i + \big(\Delta_{\mbox{\scriptsize M4}} G_\phi\big)_j{}^i \Big)\Big|_{Q=0};\qquad\\
&&\hspace*{-7mm} \Delta_{\mbox{\scriptsize B11}}\Big(\frac{\beta}{\alpha_0^2}\Big) = \frac{C_2}{\pi} \frac{d}{d\ln\Lambda}\Big( - \left(\Delta_{\mbox{\scriptsize A1}} G_c\right) \left(\Delta_{\mbox{\scriptsize A2}} G_c\right) + \Delta_{\mbox{\scriptsize A11}} G_c + \ldots\Big)\Big|_{Q=0};\\
&&\hspace*{-7mm} \Delta_{\mbox{\scriptsize B12}}\Big(\frac{\beta}{\alpha_0^2}\Big) = \frac{C_2}{\pi} \frac{d}{d\ln\Lambda} \Big( -\frac{1}{2}\left(\Delta_{\mbox{\scriptsize A2}} G_c\right)^2 + \ldots\Big)\Big|_{Q=0};\\
\label{Relation_Last}
&&\hspace*{-7mm} \Delta_{\mbox{\scriptsize B13}}\Big(\frac{\beta}{\alpha_0^2}\Big) = \frac{C_2}{\pi} \frac{d}{d\ln\Lambda} \Big(\Delta_{\mbox{\scriptsize A13}} G_c + \ldots\Big)\Big|_{Q=0}.
\end{eqnarray}

\noindent
Here the condition $Q=0$ indicates the limit of the vanishing external (Euclidean) momentum, $\Delta_{\mbox{\scriptsize B}n}(\beta/\alpha_0^2)$ denotes a contribution to the $\beta$-function generated by the graph $\mbox{B}n$, and $\Delta_{\mbox{\scriptsize A}n} G_c$ and $(\Delta_{\mbox{\scriptsize M}n} G_\phi)_j{}^i$ are the contributions to the functions $G_c$ and $(G_\phi)_j{}^i$ coming from the diagrams $\mbox{A}n$ and $\mbox{M}n$, respectively. The dots denote terms coming from the cuts of internal gauge lines. It is important that Eqs. (\ref{Relation_First}) --- (\ref{Relation_Last}) are valid at the level of loop integrals. This confirms the qualitative picture which explains appearance of Eq. (\ref{NSVZ_Equation_New_Form}) in the perturbation theory suggested in \cite{Stepanyantz:2016gtk}.

Summing up the results for all considered supergraphs we obtain the relation

\begin{eqnarray}\label{Main_Result}
&& \Delta\Big(\frac{\beta}{\alpha_0^2}\Big) = \frac{C_2}{\pi} \frac{d}{d\ln\Lambda} \Delta\ln G_c\Big|_{Q=0}  -\frac{1}{2\pi r} C(R)_i{}^j \frac{d}{d\ln\Lambda} \big(\Delta \ln G_\phi\big)_j{}^i\Big|_{Q=0} + \ldots\qquad\nonumber\\
&& = \frac{C_2}{\pi} \Delta\gamma_c  -\frac{1}{2\pi r} C(R)_i{}^j \big(\Delta\gamma_\phi\big)_j{}^i + \ldots\qquad
\end{eqnarray}

\noindent
In this equation dots denote the omitted terms which appear when internal gauge lines are cut by total derivatives. As we discussed earlier, they cannot be separately compared with $\gamma_V$. The two-loop anomalous dimension of the Faddeev--Popov ghosts $\Delta\gamma_c$ is given by the superdiagrams presented in Fig.~\ref{Figure_Ghost_Gamma_Diagrams}. In the form of loop integrals their sum is written as

\begin{eqnarray}\label{Gamma_C_Integral_Form}
&& \Delta\gamma_c = \frac{d}{d\ln\Lambda}\Bigg\{ 4\pi C_2  \int \frac{d^4K}{(2\pi)^4} \frac{\alpha_0}{K^4 R_K}\left[(\xi_0-1) \Big(\frac{1}{3} - \frac{5}{2} y_0 C_2\Big)\right.\nonumber\\
&&\qquad\qquad\qquad\qquad\qquad\qquad\qquad\qquad\qquad\qquad\  \left. + \frac{8\pi\alpha_0}{3 R_K}\Big(C_2 f(K/\Lambda)+ T(R) h(K/\Lambda)\Big) \right] \nonumber\\
&& + 4\pi^2 C_2^2 \alpha_0^2 \int \frac{d^4K}{(2\pi)^4} \frac{d^4L}{(2\pi)^4} \frac{1}{R_K R_L} \left[\frac{(\xi_0-1)(5\xi_0+8)}{9 K^4 L^4}  - \frac{4(\xi_0^2-1)}{3 K^4 L^2 (K+L)^2} \right] \Bigg\}\Bigg|_{\alpha,\xi,y=\mbox{\scriptsize const}},\qquad\quad
\end{eqnarray}

\noindent
where the expressions for the functions $f(K/\Lambda)$ and $h(K/\Lambda)$ can be found in Appendix \ref{Appendix_Functions}. The expression (\ref{Gamma_C_Integral_Form}) has been calculated in Ref. \cite{Kazantsev:2018kjx} for the regulator $R(x) = 1+x^n$, where $n\ge 1$ is a positive integer,

\begin{eqnarray}\label{Bare_Ghost_Gamma_With_Y}
&& \Delta\gamma_c = \frac{\alpha_0 C_2 (\xi_0-1)}{6\pi} - \frac{5\alpha_0 y_0 C_2^2 (\xi_0-1)}{4\pi}  -  \frac{\alpha_0^2 C_2^2}{24\pi^2} \big(\xi_0^2-1\big) - \frac{\alpha_0^2 C_2^2}{4\pi^2}\big( \ln a_\varphi + 1\big) \qquad \nonumber\\
&& + \frac{\alpha_0^2 C_2 T(R)}{12\pi^2}\big( \ln a + 1\big),
\end{eqnarray}

\noindent
where $a$ and $a_\varphi$ are the regularization parameters defined by Eq. (\ref{Pauli_Villars_Masses}). It is important that this expression contains the parameter $y_0$ which appears due to the nonlinear renormalization of the quantum gauge superfield $V$. In Eq. (\ref{Bare_Ghost_Gamma_With_Y}) the dependence on this parameter has been calculated only in the lowest one-loop approximation. Without it the renormalization group equations are not satisfied. However, in the two-loop approximation  the parameters describing the nonlinear renormalization have not been taken into account. That is why the complete two-loop result for $\gamma_c$ can be written only in the gauge $y_0=0$, which corresponds to ${\cal F}(V) = V$,

\begin{equation}\label{Bare_Ghost_Gamma}
\Delta\gamma_c\Big|_{y_0=0} = \frac{\alpha_0 C_2 (\xi_0-1)}{6\pi} -  \frac{\alpha_0^2 C_2^2}{24\pi^2} \big(\xi_0^2-1\big) - \frac{\alpha_0^2 C_2^2}{4\pi^2}\big(\ln a_\varphi + 1\big) + \frac{\alpha_0^2 C_2 T(R)}{12\pi^2}\big(\ln a + 1\big).
\end{equation}

\noindent
The expression $\big(\Delta\gamma_\phi\big)_j{}^i$ in Eq. (\ref{Main_Result}) is a part of the matter superfield anomalous dimension corresponding to the sum of the superdiagrams M1 --- M4 presented in Fig.~\ref{Figure_Matter_Anomalous_Dimension},

\begin{eqnarray}\label{Gamma_Phi}
&& \big(\Delta\gamma_\phi\big)_j{}^i = \frac{d}{d\ln\Lambda} \Big(\big(\Delta_{\mbox{\scriptsize M1}} G_\phi\big)_j{}^i + \big(\Delta_{\mbox{\scriptsize M2}} G_\phi\big)_j{}^i
+ \big(\Delta_{\mbox{\scriptsize M3}} G_\phi\big)_j{}^i + \big(\Delta_{\mbox{\scriptsize M4}} G_\phi\big)_j{}^i \Big)\Big|_{Q=0}\qquad\nonumber\\
&& = C_2 C(R)_j{}^i \frac{d}{d\ln\Lambda} \int \frac{d^4K}{(2\pi)^4} \frac{d^4L}{(2\pi)^4} \frac{e_0^4}{R_K^2}\Big(-\frac{1}{(K+L)^2 L^2 K^4} + \frac{2}{3 L^2 K^6} \Big).
\end{eqnarray}

\noindent
We see that this expression is not well-defined. However, it is quite expected, because the well-defined results are obtained only after summations of all supergraphs. In particular, to find the well-defined expression for $\big(\gamma_\phi\big)_j{}^i$, one should take into consideration all relevant superdiagrams which are obtained by cutting supergraphs with the matter loop(s).

Eq. (\ref{Main_Result}) is the main result of this paper. It demonstrates that the NSVZ relation in the form (\ref{NSVZ_Equation_New_Form}) is really valid for the supergraphs containing the ghost loop(s) in the considered order of the perturbation theory. Although the total three-loop calculation has not yet been done, the result obtained here allows to verify the term containing the anomalous dimension $\gamma_c$ in the approximation where the scheme-dependence becomes essential. (For RGFs defined in terms of the bare couplings, which are considered in this paper, this means the dependence on a regularization.)

\section*{Conclusion}
\hspace*{\parindent}

In this paper we have verified the NSVZ relation in the form (\ref{NSVZ_Equation_New_Form}) by comparing the three-loop contribution to the $\beta$-function coming from the superdiagrams containing loops of the Faddeev--Popov ghosts with the two-loop contribution to the anomalous dimension of these ghosts. The check is made in the case of using the higher covariant derivative regularization for RGFs defined in terms of the bare couplings. It is very nontrivial, because in this approximation the scheme dependence becomes essential. Moreover, in this calculation the nonlinear renormalizaton of the quantum gauge superfield is also very important.

The verification is based on the possibility of matching the superdiagrams contributing to the $\beta$-function and the superdiagrams contributing to the anomalous dimensions of the quantum superfields. The former ones are obtained from a certain vacuum supergraph by attaching two external lines of the background gauge superfield in all possible ways, while the latter ones are generated by all possible cuts of internal lines in the considered vacuum supergraph. In the case of using the higher covariant derivative regularization one can not only match various groups of superdiagrams, but also relate them by equations analogous to (\ref{NSVZ_Equation_New_Form}). This can be done, because in this case the integrals giving the $\beta$-function are integrals of double total derivatives in any order of the perturbation theory \cite{Stepanyantz:2019ihw}. It is these double total derivatives that cut internal lines with the help of the identity (\ref{Delta_Identity}). Thus, it is possible to identify what lines should be cut for obtaining various terms in the right hand side of Eq. (\ref{NSVZ_Equation_New_Form}).

In this paper we consider all two- and three-loop vacuum supergraphs containing ghost loops and compare the parts of the $\beta$-function corresponding to them with the relevant contributions to the anomalous dimensions $\gamma_c$ and $(\gamma_\phi)_i{}^j$.\footnote{The contributions to $\gamma_V$ can be compared only if the sum of the corresponding superdiagrams with two external lines of the quantum gauge superfield is transversal. This can be achieved only by taking into consideration the purely gauge supergraphs which are not investigated in this paper.} The contributions to the $\beta$-function were obtained using a special method proposed in Ref. \cite{Stepanyantz:2019ihw}, which allows to essentially simplify the calculations. This method requires calculating only (specially modified) vacuum supergraphs and produces the result in the form of an integral of double total derivatives. Then we find singular contributions which originate from cutting of the internal ghost and matter lines and compare their sums with the corresponding parts of $\gamma_c$ and $(\gamma_\phi)_i{}^j$. This comparison reveals that Eq. (\ref{NSVZ_Equation_New_Form}) is really valid for the considered terms at the level of loop integrals. Thus, the correctness of the general results discussed in Refs. \cite{Stepanyantz:2016gtk}
and \cite{Stepanyantz:2019ihw} is confirmed by a highly nontrivial calculation.

\section*{Acknowledgements}
\hspace*{\parindent}

The authors are very grateful to A.E.Kazantsev for numerous useful discussions and valuable comments on the manuscript.

This research was supported by Foundation for Advancement of Theoretical Physics and Mathematics ``BASIS'', grants No. 17-11-120-42 (M.K.),
18-2-6-159-1 (N.M.), 18-2-6-158-1 (S.N.), 19-1-1-45-3 (I.S.), and 19-1-1-45-1 (K.S.).

\appendix

\section*{Appendix}

\section{Results for the supergraphs}
\hspace*{\parindent}\label{Appendix_Ghost_Supergraphs}

In this appendix we collect the expressions for contributions of the supergraphs presented in Fig. \ref{Figure_Ghost_Beta_Diagrams} to the function $\beta/\alpha_0^2$. All of them are constructed according to the algorithm described in Sect. \ref{Section_Algorithm}. Bold letters denote the inverse squared momenta coming from the ghost and matter propagators. (The cuts of these propagators produce contributions to $\gamma_c$ and $(\gamma_\phi)_i{}^j$.)

\begin{eqnarray}
&&\hspace*{-5mm} \Delta_{\mbox{\scriptsize B1}}\Big(\frac{\beta}{\alpha_0^2}\Big) =  \pi C_2^2\, \frac{d}{d\ln\Lambda} \int \frac{d^4Q}{(2\pi)^4} \frac{d^4K}{(2\pi)^4} \Big( \frac{\partial^2 }{\partial Q_{\mu}^2} +\frac{\partial^2 }{\partial K_{\mu}^2}-\frac{\partial^2}{\partial Q_{\mu} \partial K^{\mu}} \Big)\frac{e_0^2\, (\xi_0-1)}{K^4 R_K}\nonumber\\
&&\hspace*{-5mm} \times \left\{\frac{K^2}{\bm{(K+Q)^2 Q^2}} -\frac{1}{\bm{Q^2}} -\frac{1}{\bm{(K+Q)^2}} \right\};\\
\vphantom{1}\nonumber\\
&&\hspace*{-5mm} \Delta_{\mbox{\scriptsize B2}}\Big(\frac{\beta}{\alpha_0^2}\Big) =  \frac{4\pi}{3} C_2^2\, \frac{d}{d\ln\Lambda} \int \frac{d^4Q}{(2\pi)^4} \frac{d^4K}{(2\pi)^4} \Big(\frac{\partial^2}{\partial Q_\mu^2} + \frac{\partial^2}{\partial K_\mu^2}\Big)\frac{e_0^2\, (\xi_0-1)}{K^4 R_K \bm{Q^2}}\Big(1 - \frac{15}{2} y_0 C_2\Big);\\
\vphantom{1}\nonumber\\
&&\hspace*{-5mm} \Delta_{\mbox{\scriptsize B3}}\Big(\frac{\beta}{\alpha_0^2}\Big) = \frac{\pi}{2} C_2^3 \frac{d}{d \ln \Lambda} \int \frac{d^4Q}{(2\pi)^4}\frac{d^4K}{(2\pi)^4}\frac{d^4L}{(2\pi)^4}
\Big(\frac{\partial^2}{\partial Q_{\mu}^2} + \frac{\partial^2}{\partial K_{\mu}^2} + \frac{\partial^2}{\partial L_{\mu}^2} - \frac{\partial^2}{\partial Q_{\mu} \partial K^{\mu}} - \frac{2}{3}\frac{\partial^2}{\partial Q_{\mu}\partial L^{\mu}}
\nonumber\\
&&\hspace*{-5mm}  - \frac{1}{3}\frac{\partial^2}{\partial K_{\mu} \partial L^{\mu}} \Big) \frac{e_0^4(\xi_0-1)}{R_K R_L K^2 L^2}
\left\{ -\frac{\xi_0+1}{\bm{Q^2 (K+Q+L)^2 (K+Q)^2}} + \frac{\xi_0 -1}{K^2 L^2} \Big(\frac{1}{\bm{Q^2}} + \frac{1}{\bm{(Q+K)^2}}\right.\nonumber\\
&&\hspace*{-5mm}  + \frac{1}{\bm{(Q+K+L)^2}} + \frac{(Q+K)^2}{\bm{Q^2 (Q+K+L)^2}} \Big) + \frac{1}{L^2\bm{Q}^2} \Big(\frac{1}{\bm{(Q+K+L)^2}} + \frac{1}{\bm{(Q+K)^2}}\Big)\nonumber\\
&&\hspace*{-5mm} \left.  +\frac{1}{K^2\bm{(Q+K+L)^2}} \Big(\frac{1}{\bm{Q^2}} + \frac{1}{\bm{(Q+K)^2}}\Big) \right\};\\
\vphantom{1}\nonumber\\
&&\hspace*{-5mm} \Delta_{\mbox{\scriptsize B4}} \Big(\frac{\beta}{\alpha^2_0} \Big) = -\frac{\pi}{2} C_2^3 \frac{d}{d\ln\Lambda} \int \frac{d^4Q}{(2\pi)^4} \frac{d^4K}{(2\pi)^4} \frac{d^4L}{(2\pi)^4} \Big(\frac{\partial^2}{\partial Q_\mu^2} + \frac{\partial^2}{\partial K_\mu^2} + \frac{\partial^2}{\partial L_\mu^2} - \frac{\partial^2}{\partial Q_\mu \partial L^\mu}  \nonumber\\
&&\hspace*{-5mm} - \frac{\partial^2}{\partial Q_\mu \partial K^\mu}  +\frac{1}{2} \frac{\partial^2}{\partial K_\mu \partial L^\mu} \Big) \frac{ e_0^4 (\xi_0 - 1)^2}{R_K R_L K^2 L^2} \left\{ \frac{1}{\bm{Q^2 (Q+K)^2 (Q+L)^2}} - \frac{1}{K^2 \bm{(Q+L)^2}}\Big(\frac{1}{\bm{Q^2}} \right.\nonumber\\
&&\hspace*{-5mm} + \frac{1}{\bm{(Q+K)^2}}\Big) - \frac{1}{L^2 \bm{(Q+K)^2}}\Big(\frac{1}{\bm{Q^2}} + \frac{1}{\bm{(Q+L)^2}}\Big) + \frac{1}{K^2 L^2} \Big( \frac{1}{\bm{(K+Q)^2}} + \frac{1}{\bm{(Q+L)^2}} \nonumber\\
&&\hspace*{-5mm}\left. +\frac{Q^2}{\bm{(K+Q)^2  (Q+L)^2}} + \frac{1}{\bm{Q^2}} \Big)\right\};\\
\vphantom{1}\nonumber\\
&&\hspace*{-5mm} \Delta_{\mbox{\scriptsize B5}}\Big(\frac{\beta}{\alpha_0^2}\Big) = \frac{\pi}{8} C_2^3 \frac{d}{d \ln \Lambda}  \int \frac{d^4 Q}{(2\pi)^4}\frac{d^4 K}{(2\pi)^4}\frac{d^4 L}{(2\pi)^4}\Big(\frac{\partial^2}{\partial Q_{\mu}^2}  + \frac{\partial^2}{\partial K_{\mu}^2}
+ \frac{\partial^2}{\partial L_{\mu}^2} - \frac{\partial^2}{\partial Q_\mu \partial K^\mu} - \frac{\partial^2}{\partial Q_\mu \partial L^\mu}
\Big) \nonumber\\
&&\hspace*{-5mm} \times \frac{e_0^4 (\xi_0-1)}{R_K R_L K^2 L^2} \left\{
- \frac{1}{K^2} \Big(\frac{1}{\bm{Q^2 (Q+L)^2}} + \frac{1}{\bm{(Q+K)^2 (Q+K+L)^2}}\Big) - \frac{1}{L^2} \Big(\frac{1}{\bm{Q^2 (Q+K)^2}}\right.\nonumber\\
&&\hspace*{-5mm} + \frac{1}{\bm{(Q+L)^2 (Q+K+L)^2}}\Big) - \frac{\xi_0(K^2+L^2)}{K^2L^2}\Big(\frac{1}{\bm{Q^2(Q+K+L)^2}}+\frac{1}{\bm{(Q+K)^2 (Q+L)^2}}\Big)\nonumber\\
&&\hspace*{-5mm} + \frac{(\xi_0+1) (2Q+K+L)^2 + \xi_0 (L^2 + K^2)}{\bm{Q^2 (Q+K)^2 (Q+L)^2 (K+Q+L)^2}} -  \frac{(\xi_0 -1)}{K^2 L^2} \Big(\frac{1}{\bm{Q^2}} + \frac{1}{\bm{(Q+K)^2}} + \frac{1}{\bm{(Q+L)^2}}  \nonumber\\
&&\hspace*{-5mm} \left. +\frac{1}{\bm{(Q+K+L)^2}}\Big)\right\};\\
&&\hspace*{-5mm} \vphantom{1}\nonumber\\
&&\hspace*{-5mm}\Delta_{\mbox{\scriptsize B6}} \Big(\frac{\beta}{\alpha^2_0}\Big)  = - \frac{\pi}{6} C_2^3 \frac{d}{d\ln\Lambda} \int \frac{d^4 Q}{\left(2 \pi \right)^4} \frac{d^4 K}{\left(2 \pi \right)^4} \frac{d^4 L}{\left(2 \pi \right)^4} \Big(\frac{\partial^2}{\partial Q_\mu^2} +  \frac{\partial^2}{\partial K_\mu^2} + \frac{\partial^2}{\partial L_\mu^2} - \frac{\partial^2}{\partial Q_\mu \partial K^\mu} -\frac{2}{3}\frac{\partial^2}{\partial Q_\mu \partial L^\mu}\nonumber \\
&&\hspace*{-5mm}   - \frac{1}{3} \frac{\partial^2}{\partial K_\mu \partial L^\mu} \Big) \frac{e_0^4 (\xi_0 - 1)}{R_K R_L K^4 L^4} \left\{ \frac{2K^2 + 2L^2}{\bm{Q^2 (Q+K+L)^2}} + \frac{1}{2}(\xi_0 - 1)\Big( \frac{1}{\bm{Q^2}} + \frac{1}{\bm{(Q+K+L)^2}}  \right.\nonumber\\
&&\hspace*{-5mm} \left. +\frac{(K+L)^2 + K^2 + L^2 + (Q+K)^2+ (Q+L)^2}{\bm{Q^2 (Q+K+L)^2}} \Big) \right\};\\
&&\hspace*{-5mm} \vphantom{1}\nonumber\\
&&\hspace*{-5mm} \Delta_{\mbox{\scriptsize B7}}\Big(\frac{\beta}{\alpha_0^2}\Big) = 0;\\
&&\hspace*{-5mm} \vphantom{1}\nonumber\\
&&\hspace*{-5mm} \Delta_{\mbox{\scriptsize B8}} \Big(\frac{\beta}{\alpha_0^2}\Big) =\frac{\pi}{2} C_2^3 \frac{d}{d \ln \Lambda}\int \frac{d^4 Q}{(2\pi)^4}\frac{d^4 K}{(2\pi)^4}\frac{d^4 L}{(2\pi)^4}\Big(\frac{\partial^2}{\partial Q_\mu^2}+\frac{\partial^2}{\partial K_\mu^2}+\frac{\partial^2}{\partial L_\mu^2}+\frac{\partial^2}{\partial Q^\mu\partial K_\mu}- \frac{\partial^2}{\partial Q^\mu\partial L_\mu}
\nonumber\\
&&\hspace*{-5mm} -\frac{\partial^2}{\partial K^\mu\partial L_\mu}\Big) \frac{e_0^4}{R_K K^2 R_L L^2 (K+L)^2} \left\{ (\xi_0^2-1)\Big(-\frac{2}{\bm{Q^2(Q+L)^2}} + \frac{(K+L)^2}{\bm{Q^2(Q-K)^2(Q+L)^2}}\Big)\right.\nonumber\\
&&\hspace*{-5mm} + \frac{2(\xi_0-1)}{K^2}\Big(\frac{(Q-K)^2+L^2-(K+L)^2}{\bm{Q^2(Q+L)^2}}-\frac{1}{\bm{Q^2}}\Big)
+\frac{(\xi_0-1)^2(K+L)^2}{K^2L^2} \Big( \frac{Q^2}{\bm{(Q+L)^2(Q-K)^2}}\nonumber\\
&&\hspace*{-5mm} -\frac{1}{\bm{Q^2}}\Big) +\frac{2(R_L-R_K)}{(L^2-K^2)R_{K+L}}
\left[\frac{L^2}{\bm{Q^2 (Q+L)^2}}\Big(1 -\frac{K^2}{\bm{(Q-K)^2}} \Big) - \frac{Q^2+(K+L)^2-2L^2}{\bm{(Q-K)^2(Q+L)^2}} +\frac{1}{\bm{Q^2}}\right.\nonumber\\
&&\hspace*{-5mm} \left.\left.  -\frac{(\xi_0-1)(L^2-K^2)K^2}{(K+L)^2\bm{Q^2(Q-K)^2}} \right] \right\};\\
&&\hspace*{-5mm} \vphantom{1}\nonumber\\
\label{B9}
&&\hspace*{-5mm} \Delta_{\mbox{\scriptsize B9}}\Big( \frac{\beta}{\alpha_{0}^2} \Big)= 2\pi C_2 \frac{d}{d \ln \Lambda} \int \frac{d^4 Q}{(2\pi^4)} \frac{d^4 K}{(2\pi^4)} \frac{d^4 L}{(2\pi^4)}\left[ C_{2}^2 \Big( \frac{\partial^2}{\partial K_{\mu}^2} +\frac{\partial^2}{\partial Q_{\mu}^2} +\frac{\partial^2}{\partial L_{\mu}^2}+ \frac{1}{2}\frac{\partial^2}{\partial Q_{\mu} \partial L^{\mu}} \right. \nonumber\\
&&\hspace*{-5mm}  -\frac{\partial^2}{\partial Q_{\mu} \partial K^{\mu}}  -\frac{\partial^2}{\partial L_{\mu} \partial K^{\mu}}
\Big) \frac{e_{0}^4}{R_{K}^2 K^4} \left\{\frac{K^2}{(K+Q)^2 Q^2} -\frac{1}{Q^2} -\frac{1}{(K+Q)^2} \right\} \left( f(K,L)+g(\xi_{0},K,L) \vphantom{\frac{1}{2}}\right. \nonumber\\
&&\hspace*{-5mm} \left. - \frac{(\xi_0^2-1)}{8L^2 (K+L)^2} + \frac{(\xi_0^2-1)}{8 K^2 (K+L)^2} - \frac{(\xi_0^2-1)}{24 K^2 L^2} \right) + \left\{ C_{2} T(R) \Big(\frac{\partial^2 }{\partial K_{\mu}^2}+\frac{\partial^2 }{\partial Q_{\mu}^2} + \frac{1}{2}\frac{\partial^2}{\partial Q_{\mu} \partial L^{\mu}}  \right.\nonumber\\
&&\hspace*{-5mm} \left.-\frac{\partial^2}{\partial Q_{\mu} \partial K^{\mu}} -\frac{\partial^2}{\partial L_{\mu} \partial K^{\mu}}\Big) + \frac{1}{r} \mbox{tr}\left[C(R)^2\right] \frac{\partial^2}{\partial L_{\mu}^2} \right\}  \left. \frac{e_{0}^4}{R_{K}^2 K^4}\left\{ \frac{K^2}{(K+Q)^2 Q^2} -\frac{1}{Q^2} -\frac{1}{(K+Q)^2} \right\}\right.\nonumber\\
&&\hspace*{-5mm} \left.\vphantom{\frac{1}{2}} \times h(K,L) \right];\\
\vphantom{1}\nonumber\\
\label{B10}
&&\hspace*{-5mm} \Delta_{\mbox{\scriptsize B10}}\Big( \frac{\beta}{\alpha_{0}^2} \Big)= \frac{8\pi}{3} C_2 \frac{d}{d \ln \Lambda} \int \frac{d^4 Q}{(2\pi^4)} \frac{d^4 K}{(2\pi^4)} \frac{d^4 L}{(2\pi^4)}
\left[ C_{2}^2 \Big(\frac{\partial^2}{\partial K_{\mu}^2} +\frac{\partial^2}{\partial Q_{\mu}^2} +\frac{\partial^2}{\partial L_{\mu}^2}+ \frac{1}{2}\frac{\partial^2}{\partial Q_{\mu} \partial L^{\mu}}\right.
\nonumber\\
&&\hspace*{-5mm} -\frac{\partial^2}{\partial Q_{\mu} \partial K^{\mu}}  -\frac{\partial^2}{\partial L_{\mu} \partial K^{\mu}} \Big)\frac{e_0^4}{R_K^2 K^4 Q^2} \left( f(K,L)+g(\xi_{0},K,L)
- \frac{(\xi_0^2-1)}{8 L^2 (K+L)^2} + \frac{(\xi_0^2-1)}{8 K^2 (K+L)^2}\right. \nonumber\\
&&\hspace*{-5mm} \left. - \frac{(\xi_0^2-1)}{24 K^2 L^2}\right) + \vphantom{\left\{\frac{1}{2}\right\}}\left\{ C_{2} T(R) \Big(\frac{\partial^2 }{\partial K_{\mu}^2}+\frac{\partial^2 }{\partial Q_{\mu}^2} + \frac{1}{2}\frac{\partial^2}{\partial Q_{\mu} \partial L^{\mu}} -\frac{\partial^2}{\partial Q_{\mu} \partial K^{\mu}} -\frac{\partial^2}{\partial L_{\mu} \partial K^{\mu}}\Big)\right. \nonumber\\
&&\hspace*{-5mm} \left.\left. + \frac{1}{r} \mbox{tr}\left[C(R)^2\right] \frac{\partial^2}{\partial L_{\mu}^2} \right\} \frac{e_{0}^4}{R_{K}^2 K^4 Q^2} h(K,L) \right];\\
\vphantom{1}\nonumber\\
&&\hspace*{-5mm} \Delta_{\mbox{\scriptsize B11}} \Big(\frac{\beta}{\alpha^2_0} \Big)  = - \frac{2\pi}{3} C_2^3 \frac{d}{d\ln\Lambda} \int \frac{d^4Q}{(2\pi)^4} \frac{d^4K}{(2\pi)^4} \frac{d^4L}{(2\pi)^4} \left\{\frac{\partial^2}{\partial Q_\mu^2} + \frac{\partial^2}{\partial K_\mu^2} + \frac{\partial^2}{\partial L_\mu^2}  - \frac{\partial^2}{\partial Q_\mu \partial K^\mu} \right\}\nonumber \\
&&\hspace*{-5mm} \times \frac{e_0^4 (\xi_0 - 1)^2}{R_K R_L K^4 L^4} \left\{ \frac{K^2}{\bm{Q^2 (K+Q)^2}} - \frac{1}{\bm{Q^2}} - \frac{1}{\bm{(K+Q)^2}} \right\};\\
\vphantom{1}\nonumber\\
&&\hspace*{-5mm} \Delta_{\mbox{\scriptsize B12}}\Big(\frac{\beta}{\alpha_0^2}\Big) =  -\frac{2\pi}{9} C_2^3 \frac{d}{d\ln\Lambda} \int \frac{d^4Q}{(2\pi)^4} \frac{d^4K}{(2\pi)^4} \frac{d^4L}{(2\pi)^4} \left(\frac{\partial^2}{\partial Q_\mu^2} + 2\frac{\partial^2}{\partial K_\mu^2}\right)\frac{e_0^4 (\xi_0-1)^2}{R_K R_L K^4 L^4 \bm{Q^2}};\\
\vphantom{1}\nonumber\\
&&\hspace*{-5mm} \Delta_{\mbox{\scriptsize B13}}\Big(\frac{\beta}{\alpha_0^2}\Big) =  -\frac{2\pi}{9} C_2^3 \frac{d}{d\ln\Lambda} \int \frac{d^4Q}{(2\pi)^4} \frac{d^4K}{(2\pi)^4} \frac{d^4L}{(2\pi)^4} \left(\frac{\partial^2}{\partial Q_\mu^2} + 2\frac{\partial^2}{\partial K_\mu^2}\right)\frac{e_0^4 (\xi_0-1)^2}{R_K R_L K^4 L^4 \bm{Q^2}}.
\end{eqnarray}

\noindent
The explicit form of the functions $f(K,L)$, $g(\xi_0,K,L)$, and $h(K,L)$ (which are present in Eqs. (\ref{B9}) and (\ref{B10})) can be found in Appendix \ref{Appendix_Functions}.

To obtain Eqs. (\ref{B9}) and (\ref{B10}), we first calculate the effective diagrams B9 and B10 and, after this, subtract from the result the expressions for the corresponding supergraphs with two ghost loops. (One of them includes a propagator with the momentum $Q^\mu$, and the other includes a propagator with the momentum $L^\mu$). The subtracted expressions correspond to the terms proportional to $\xi_0^2-1$ in the round brackets. This allows to avoid the double summation and take into account the factor $1/2$ needed for these supergraphs. However, in resulting expressions it is very difficult to separate the momenta corresponding to gauge and ghost internal lines. That is why in Eqs. (\ref{B9}) and (\ref{B10}) we do not use bold letters. Nevertheless, it is possible to rewrite the sum of these expressions in a different form. For this purpose we first note that no propagators in the considered supergraphs contain $Q^\mu$ and $L^\mu$ together. Therefore, the derivative $\partial^2/\partial Q_\mu \partial L^\mu$ does not contribute to the result. Also it is easy to see that due to the identities

\begin{equation}
\int \frac{d^4Q}{(2\pi)^4} \frac{\partial^2}{\partial K_\mu^2} \Big(\frac{K^2}{\bm{Q^2 (Q+K)^2}} - \frac{1}{\bm{Q^2}} - \frac{1}{\bm{(Q+K)^2}}\Big) = 0;\qquad \int \frac{d^4Q}{(2\pi)^4} \frac{\partial^2}{\partial K_\mu^2} \frac{1}{\bm{Q^2}} = 0
\end{equation}

\noindent
the derivative $\partial^2/\partial K_\mu^2$ produces only the cuts of internal gauge lines and, therefore, is also not essential for this calculation. Next, we note that in the sum of the supergraphs with two ghost loops the ghost momenta $Q^\mu$ and $L^\mu$ enter symmetrically. This implies that omitting the terms containing the derivatives with respect to $L^\mu$ we effectively divide the result by 2. (Consequently, in this case the superdiagrams with two ghost loops should not be subtracted.) In the supergraphs with a single ghost loop and no matter loops we denote the momentum of the ghost loop by $Q^\mu$, so that the derivatives $\partial/\partial L^\mu$ will cut only internal gauge lines and can also be omitted. Taking into account that the supergraphs with a matter loop correspond to the terms containing the function $h(K,L)$, the contribution of B9 and B10 to the function $\beta/\alpha_0^2$ can be equivalently rewritten in the form

\begin{eqnarray}
&&\hspace*{-5mm} \Delta_{\mbox{\scriptsize B9}}\Big( \frac{\beta}{\alpha_{0}^2} \Big) + \Delta_{\mbox{\scriptsize B10}}\Big( \frac{\beta}{\alpha_{0}^2} \Big) = 2\pi C_2 \frac{d}{d \ln \Lambda} \int \frac{d^4 Q}{(2\pi)^4} \frac{d^4 K}{(2\pi)^4} \frac{d^4 L}{(2\pi)^4}\left[ C_{2}^2 \Big(\frac{\partial^2 }{\partial Q_{\mu}^2}-\frac{\partial^2}{\partial Q_{\mu} \partial K^{\mu}} \Big) \frac{e_{0}^4}{R_{K}^2 K^4} \right. \nonumber\\
&&\hspace*{-5mm} \times  \left\{\frac{K^2}{\bm{(K+Q)^2 Q^2}} +\frac{1}{3\bm{Q^2}} -\frac{1}{\bm{(Q+K)^2}}\right\} \Big (f(K,L)+g(\xi_{0},K,L) \Big)+ \left\{ C_{2} T(R) \Big(\frac{\partial^2 }{\partial K_{\mu}^2}+\frac{\partial^2}{\partial Q_{\mu}^2}\right.\nonumber\\
&&\hspace*{-5mm} \left. -\frac{\partial^2}{\partial Q_{\mu} \partial K^{\mu}} -\frac{\partial^2}{\partial L_{\mu} \partial K^{\mu}} \Big)+ \frac{1}{r} \mbox{tr}\left[C(R)^2\right] \frac{\partial^2}{\partial L_{\mu}^2} \right\}  \left. \frac{e_{0}^4}{R_{K}^2 K^4}\left\{ \frac{K^2}{\bm{(K+Q)^2 Q^2}} +\frac{1}{3\bm{Q^2}} -\frac{1}{\bm{(Q+K)^2}}\right\}\right.
\nonumber\\
&&\hspace*{-5mm} \times\left.\vphantom{\frac{1}{2}} h(K,\bm{L}) \right] + \mbox{terms giving the cuts of internal gauge lines only}.\vphantom{\frac{1}{2}}
\end{eqnarray}

\noindent
Here the gauge and ghost internal momenta are separated. This allows to denote the inverse squared ghost and matter momenta by the bold letters as in the expressions for the other supergraphs.

The expressions for $\Delta_{\mbox{\scriptsize B}n}(\beta/\alpha_0^2)$ should be compared with the corresponding contributions to the ghost and matter anomalous dimensions, see Eqs. (\ref{Relation_First}) --- (\ref{Relation_Last}). For completeness, here we also present the contributions to $G_c$ calculated in Ref. \cite{Kazantsev:2018kjx}. Some of them vanish,

\begin{eqnarray}
&& \Delta_{\mbox{\scriptsize A1}} G_c\Big|_{Q=0} = 0;\qquad\ \Delta_{\mbox{\scriptsize A3}} G_c\Big|_{Q=0} = 0;\qquad\ \Delta_{\mbox{\scriptsize A4}} G_c\Big|_{Q=0} = 0;\qquad\ \Delta_{\mbox{\scriptsize A5}} G_c\Big|_{Q=0} = 0; \nonumber\\
&& \Delta_{\mbox{\scriptsize A6}} G_c\Big|_{Q=0} = 0;\qquad\ \Delta_{\mbox{\scriptsize A7}} G_c\Big|_{Q=0} = 0;\qquad\ \Delta_{\mbox{\scriptsize A9}} G_c\Big|_{Q=0} = 0;\qquad\ \Delta_{\mbox{\scriptsize A10}} G_c\Big|_{Q=0} = 0;\qquad\nonumber\\
&& \Delta_{\mbox{\scriptsize A11}} G_c\Big|_{Q=0} = 0;\qquad \Delta_{\mbox{\scriptsize A14}} G_c\Big|_{Q=0} = 0.
\end{eqnarray}

\noindent
The nontrivial contributions are

\begin{eqnarray}\label{Diagram1}
&&\hspace*{-5mm}  \Delta_{\mbox{\scriptsize A2}} G_c\Big|_{Q=0} = e_0^2 C_2 (\xi_0-1) \Big(\frac{1}{3} - \frac{5}{2}y_0 C_2 \Big)\int \frac{d^4K}{(2\pi)^4} \frac{1}{K^4 R_K};\\
&&\hspace*{-5mm}  \Delta_{\mbox{\scriptsize A8}} G_c\Big|_{Q=0} = \frac{e_0^4 C_2^2}{4} \xi_0(\xi_0-1) \int \frac{d^4K}{(2\pi)^4} \frac{d^4L}{(2\pi)^4} \frac{1}{K^4 R_K L^4 R_L};\qquad\\
\label{Diagram12}
&&\hspace*{-5mm}  \Delta_{\mbox{\scriptsize A12}} G_c\Big|_{Q=0} = -\frac{e_0^4 C_2^2}{6} (\xi_0-1) \int \frac{d^4K}{(2\pi)^4} \frac{d^4L}{(2\pi)^4} \frac{1}{R_K R_L} \left( \frac{\xi_0+1}{K^2 L^4 (K+L)^2} + \frac{\xi_0-1}{2 K^4 L^4}\right);\qquad\\
\label{Diagram13}
&&\hspace*{-5mm}  \Delta_{\mbox{\scriptsize A13}} G_c\Big|_{Q=0} = -\frac{e_0^4 C_2^2}{18} (\xi_0-1)^2 \int \frac{d^4K}{(2\pi)^4} \frac{d^4L}{(2\pi)^4} \frac{1}{K^4 R_K L^4 R_L};\\
\label{Diagram15}
&&\hspace*{-5mm}  \Delta_{\mbox{\scriptsize A15}} G_c\Big|_{Q=0} = \frac{2e_0^4 C_2}{3} \int \frac{d^4K}{(2\pi)^4} \frac{1}{K^4 R_K^2}\Big(C_2 f(K/\Lambda) + C_2 g(\xi_0,K/\Lambda) + T(R) h(K/\Lambda)\Big).
\end{eqnarray}

The contribution of the superdiagrams M1 --- M4 to the function $\big(\gamma_\phi\big)_i{}^j$ is given by Eq. (\ref{Gamma_Phi}).

\section{Explicit expressions for the functions $h$, $g$, and $f$.}
\hspace*{\parindent}\label{Appendix_Functions}

Here we present the explicit expressions for the functions $h$, $g$, and $f$ entering Eqs. (\ref{B9}) and (\ref{B10}),

\begin{eqnarray}\label{H_Function}
&& h(K,L) \equiv \frac{1}{2L^2(L+K)^2} + \frac{1}{\big((K+L)^2-L^2\big)}\left(\vphantom{\frac{1}{2}}\right. - \frac{M^2 F'_{K+L}}{\Lambda^2 F_{K+L} \big((K+L)^2 F_{K+L}^2 + M^2\big)} \qquad \nonumber\\
&& + \frac{M^2 F'_{L}}{\Lambda^2 F_L \big(L^2 F_{L}^2 + M^2\big)} + \frac{F_{K+L}^2}{2\big((K+L)^2 F_{K+L}^2 + M^2\big)} - \frac{F_L^2}{2\big(L^2 F_L^2+M^2\big)} \left.\vphantom{\frac{1}{2}}\right);\\
\label{G_Function}
&& g(\xi_0,K,L)\equiv \frac{(\xi_0-1)}{2L^4 R_L} \Big(R_{K+L} - \frac{2}{3} R_K\Big) - \frac{(\xi_0-1)}{2 L^2 R_L (K+L)^4 R_{K+L}}\qquad\nonumber\\
&&\qquad\qquad\qquad\qquad\qquad \times (K_\mu R_K + L_\mu R_L)^2 - \frac{(\xi_0-1)^2 K^2 R_K^2}{4 L^4 R_L (K+L)^4 R_{K+L}} L^\mu (K+L)_\mu;\qquad\quad\\
\label{F_Function}
&& f(K,L) \equiv - \frac{3}{2} \left(\frac{1}{L^2(L+K)^2} -\frac{1}{(L^2+M_{\varphi}^2)((L+K)^2+M_{\varphi}^2)}\right) + \frac{R_{L}-R_{K}}{R_{L}L^2}\nonumber\\
&& \times \left(\frac{1}{(L+K)^2} -\frac{1}{L^2-K^2}\right) +\frac{2}{R_L \big((L+K)^2-L^2\big)}\left(\frac{R_{L+K}-R_{L}}{(L+K)^2-L^2} -\frac{R'_{L}}{\Lambda^2}\right) \nonumber\\
&& - \frac{1}{R_{L}R_{L+K}} \left(\frac{R_{L+K}-R_{L}}{(L+K)^2-L^2}\right)^2   - \frac{2 R_{K}K^2}{L^2(L+K)^2R_{L}R_{L+K}} \left(\frac{R_{L+K}-R_{K}}{(L+K)^2-K^2}\right)\nonumber\\
&& - \frac{L_\mu K^\mu R_{K}}{L^2 R_{L} (L+K)^2 R_{L+K}} \left(\frac{R_{L+K}-R_{L}}{(L+K)^2-L^2}\right) + \frac{2 L_\mu K^\mu}{L^2R_{L}R_{L+K}} \left(\frac{R_{L+K}-R_{K}}{(L+K)^2-K^2}\right)\nonumber\\
&&\times \left(\frac{R_{L+K}-R_{L}}{(L+K)^2-L^2}\right) - \frac{2 K^2}{(L+K)^2 R_{L} R_{L+K}} \left(\frac{R_{L}-R_{K}}{L^2-K^2}\right)^2 -\frac{K^2 L_\mu (L+K)^\mu}{L^2 (L+K)^2 R_{L} R_{L+K}} \nonumber\\
&&\times \left(\frac{R_{L}-R_{K}}{L^2-K^2}\right) \left(\frac{R_{L+K}-R_{K}}{(L+K)^2-K^2}\right) +\frac{2K^2}{\big((L+K)^2-K^2\big) L^2 R_L} \left(\frac{R_{L+K}-R_{K}}{(L+K)^2-K^2} -\frac{R'_{K}}{\Lambda^2}\right)\nonumber\\
&& -\frac{2 L_\mu K^\mu}{L^2R_{L}} \left(\frac{R_{L}}{\left(L^2-(L+K)^2\right) \left(L^2-K^2\right)}+\frac{R_{L+K}}{\left((L+K)^2-L^2\right) \left((L+K)^2-K^2\right)}\right.\nonumber\\
&& \left.+\frac{R_{K}}{\left(K^2-L^2\right)\left(K^2-(L+K)^2\right)}\right) - \frac{1}{2\big((L+K)^2-L^2\big)}\left(\frac{2 R_{L+K} R'_{L+K} (L+K)^2}{\Lambda^2\big((L+K)^2 R_{L+K}^2+M_\varphi^2\big)} \right.\nonumber\\
&& -\frac{2 R_L R'_L L^2}{\Lambda^2\big(L^2 R_L^2 + M_\varphi^2\big)} - \frac{1}{(L+K)^2 + M_\varphi^2} + \frac{1}{L^2+M_\varphi^2} +\frac{R_{L+K}^2}{(L+K)^2 R_{L+K}^2 + M_\varphi^2}\nonumber\\
&& \left. - \frac{R_L^2}{L^2 R_L^2+M_\varphi^2} \right),
\end{eqnarray}

\noindent
where

\begin{equation}
R_L' \equiv \frac{\partial R(L^2/\Lambda^2)}{\partial(L^2/\Lambda^2)};\qquad F_L' \equiv \frac{\partial F(L^2/\Lambda^2)}{\partial(L^2/\Lambda^2)}.
\end{equation}

\noindent
They are related to the corresponding functions introduced in Ref. \cite{Kazantsev:2017fdc} by the equations

\begin{eqnarray}
&& h(K/\Lambda) = \int \frac{d^4L}{(2\pi)^4}\, h(K,L);\qquad f(K/\Lambda) = \int \frac{d^4L}{(2\pi)^4}\, f(K,L);\qquad\nonumber\\
&&\qquad\qquad\qquad\ g(\xi_0,K/\Lambda) = \int \frac{d^4L}{(2\pi)^4}\, g(\xi_0,K,L);.
\end{eqnarray}

\end{document}